\documentclass[preprint,12pt]{elsarticle}

\usepackage{geometry}
\usepackage{amsmath}
\usepackage{placeins}
\usepackage{subcaption}
\usepackage{caption}
\usepackage{graphicx}
\usepackage[hyphens]{url}
\usepackage[hidelinks]{hyperref}
\usepackage{booktabs}
\usepackage{threeparttable}
\usepackage{multirow}
\usepackage{float}
\usepackage{hyperref}
\usepackage{lineno}

\usepackage{amssymb}
\usepackage{amsthm}

\usepackage{graphicx}
\usepackage{amssymb}

\usepackage{lineno}

\usepackage[dvipsnames]{xcolor}
\usepackage{ulem}

\definecolor{nyupurple}{RGB}{150, 35, 140}

\newcommand{\xzhangadd}[1]{\textcolor{black}{#1}}

\journal{arXiv}

\begin{document}

\begin{frontmatter}


\title{\textbf{The Short-term Impact of Congestion Taxes on Ridesourcing Demand and Traffic Congestion: Evidence from Chicago}}





\author[label1]{Yuan Liang}
\address[label1]{Department of Urban and Rural Planning, School of Architecture, Southwest Jiaotong University, Chengdu, China }
\ead{yuanliang@my.swjtu.edu.cn}  

\author[label1]{Bingjie Yu}
\ead{bjyu@my.swjtu.edu.cn}   

\author[label2]{Xiaojian Zhang}
\address[label2]{Department of Civil and Coastal Engineering, University of Florida}
\ead{xiaojianzhang@ufl.edu}

\author[label3]{Yi Lu}
\address[label3]{Department of Architecture and Civil Engineering, City University of Hong Kong, Hong Kong, China}
\ead{yilu24@cityu.edu.hk}

\author[label1]{Linchuan Yang\corref{cor1}}
\ead{yanglc0125@swjtu.edu.cn}     

\cortext[cor1]{Corresponding author. Postal address: School of Architecture, Southwest Jiaotong University, Hi-Tech Industrial Development Zone, Chengdu 611756, China}

\begin{abstract}
Ridesourcing is popular in many cities. Despite its theoretical benefits, a large body of studies have claimed that ridesourcing also brings (negative) externalities (e.g., inducing trips and aggravating traffic congestion). Therefore, many cities are planning to enact or have already enacted policies to regulate its use. However, these policies' effectiveness or impact on ridesourcing demand and traffic congestion is uncertain. To this end, this study applies difference-in-differences (i.e., a regression-based causal inference approach) to empirically evaluate the effects of the congestion tax policy on ridesourcing demand and traffic congestion in Chicago. It shows that this congestion tax policy significantly curtails overall ridesourcing demand but marginally alleviates traffic congestion. The results are robust to the choice of time windows and data sets, additional control variables, alternative model specifications, alternative control groups, and alternative modeling approaches (i.e., regression discontinuity in time). Moreover, considerable heterogeneity exists. For example, the policy notably reduces ridesourcing demand with short travel distances, but such an impact is gradually attenuated as the distance increases.
\end{abstract}


\end{frontmatter}



\section{Introduction}
\label{S:Introducation}
Ridesourcing, as an emerging mode of shared mobility, has experienced rapid growth worldwide in the past decade. In 2021, the global market value of the ridesourcing industry was estimated at 117 billion US dollars\footnote{\url{https://www.statista.com/statistics/1155981/ride-sharing-market-size-worldwide/}}. By providing travelers with a convenient and cost-effective travel option, ridesourcing has become popular among the general public. For example, the Uber app had 93 million users on a monthly basis\footnote{\url{https://www.statista.com/statistics/833743/us-users-ride-sharing-services/}} and generated 1.4 billion trips worldwide in the fourth quarter of 2020. The number was considerably higher before the COVID-19 pandemic, with worldwide ridership reaching 1.9 billion in the fourth quarter of 2019\footnote{\url{https://www.statista.com/statistics/946298/uber-ridership-worldwide/}}. Along with the burgeoning development of the ridesourcing industry, there is a heated debate over its impact on urban mobility.

Theoretically, ridesourcing presents various benefits in enhancing the efficiency, accessibility, and sustainability of urban mobility  \citep{jin2018ridesourcing}. First, ridesourcing may help reduce private car dependency because it provides an on-demand, door-to-door, accessible, and affordable alternative to private cars. Second, ridesourcing has a promising potential to curtail vehicle miles traveled (VMT), energy use, and vehicle fleet size required, as well as improve vehicle occupancy. By deploying large-scale efficient dispatching algorithms, ridesourcing can reduce idle miles (those traveled with no passengers), which is a major concern for on-demand travel services. Meanwhile, ridesplitting, a special version of ridesourcing, can enable travelers to share their rides by dynamically matching them with overlapping trip routes, thereby increasing car occupancy and reducing the overall VMT. Third, ridesourcing may have complementary and supplementary effects on public transit. In favor of the former, ridesourcing is able to cope with the first/last mile problems in public transit by providing convenient connections to stations. In favor of the latter, ridesourcing can serve the time and place with poor public transit services (e.g., in the suburban area and at midnight), thereby improving urban mobility \citep{wang2019ridesourcing,zhu2022comprehensive,dean2021spatial,shaheen2016mobility}

However, blindly equating the theoretical benefits and \xzhangadd{actual impacts} of ridesourcing may be deceptive. Whether these theoretical benefits can be translated into practical fruits is an enduring question, as ridesourcing is often regarded as a double-edged sword for urban mobility \citep{jin2018ridesourcing, tirachini2020ride}. Critics claim that ridesourcing can render traditional travel modes (e.g., public transit and active travel) less competitive because of its convenience and flexibility, as well as the lack of regulation. Moreover, ridesourcing has the potential to increase VMT per passenger trip via deadheading and induce additional trips \citep{tirachini2020does}. \cite{choi2022empirical} find that the extra 0.6 percent average annual growth in VMT in the Atlanta region can be attributed to ridesourcing,  \xzhangadd{while \cite{henao2019impact} reveal that ridesourcing leads to an increase in VMT by 83.5\% in the Denver region}.  \cite{wu2021assessing} estimate that ridesourcing lead to a net increase in daily VMT by 7.8 million in the US. Consequently, ridesourcing may exacerbate traffic congestion and hamper the sustainability of urban mobility. Against this background, the real role of ridesourcing in a city and its impact on urban mobility, particularly traffic congestion, have attracted considerable scholarly attention and have been constantly investigated in the last few years. Although there is a large body of related studies, they have reached mixed, even conflicting, conclusions. Regarding traffic congestion, \cite{erhardt2019transportation} and \cite{diao2021impacts} find that the emergence of ridesourcing significantly increases traffic congestion in terms of both intensity and duration. \cite{tarduno2021congestion} reveals that the emergence of ridesourcing modestly intensifies traffic congestion, which is supported by the empirical evidence that traffic congestion was relieved during the abrupt exit of Uber in Austin. Moreover, such negative effects on traffic congestion could be heterogeneous across contextual factors (e.g., urban areas with different built environments and weekdays/weekends) or even become positive. \cite{li2022demand} indicate that ridesourcing increases traffic congestion in compact urban areas but relieves traffic congestion in sprawling areas. \cite{dhanorkar2022heterogeneous} find that the crowding effects of ridesourcing on traffic congestion are more prominent on interior roads and in high-density areas. Their empirical results also show that ridesourcing decreases traffic congestion on weekdays but increases traffic congestion on weekends. \xzhangadd{In terms of car ownership, \cite{ward2021impact} identify that ridesourcing entry into urban areas, on average, contributes to a 0.7\% increase in vehicle registrations in the US, whereas \cite{diao2021impacts} document that such effects are insignificant}. As for public transit, \cite{hall2018uber} demonstrate that the entry of Uber increases the ridership of public transit across US metropolitan areas. By contrast, \cite{diao2021impacts} and \cite{erhardt2022transportation} reach an opposite conclusion. \cite{diao2021impacts} observe that ridesourcing lowers transit ridership by 8.9\%. \cite{erhardt2022transportation} find that ridesourcing reduces bus ridership by 10\% in San Francisco.

Despite the uncertain impact of ridesourcing on urban mobility, the rapid growth of ridesourcing trips is tangible and dramatic. In response to the \xzhangadd{rapid growth} of ridesourcing and its potential negative externalities\footnote{\xzhangadd{The rationale for taxing ridesourcing services also includes covering increasing regulatory costs and avoiding revenue shortfalls for key transportation priorities (e.g., supporting transit services and accessible vehicles) resulting from the shift from taxis to ridesourcing, as taxis are subject to many special surcharges and accessible vehicle requirements that ridesourcing have escaped.}} (\xzhangadd{e.g.,} traffic congestion), several municipal authorities are planning to enact or have already enacted congestion taxes to regulate ridesourcing, especially in the downtown area \citep{zhao2020revenue,lehe2021taxation}. For instance, New York started to collect a per-trip surcharge for ridesourcing trips that start or end within the downtown congestion zone on January 1$^{st}$, 2019. Chicago has adopted a new ridesourcing-targeted congestion tax (henceforth congestion tax) since January 6$^{th}$, 2020, which imposes differential per-trip surcharges based on whether the trip is shared and starts/ends within the downtown area. According to microeconomics theory, higher prices lead to less usage and consumption \citep{mankiw2020principles}. Therefore, the congestion tax is expected to reduce overall ridesourcing demand, encourage travelers to share their rides, and combat the rampant traffic congestion.

Accordingly, it is urgent and imperative to conduct rigorous ex-post evaluations on the impact of such congestion taxes on ridesourcing demand, traffic congestion, and other transportation system performance. Precise ex-post evaluations would help policymakers (1) ascertain whether the congestion taxes for ridesourcing achieve their goals and (2) test the effectiveness of such taxes. Moreover, it can provide valuable experiences and references for the remedies and future decision-making process. Such evaluations also contribute to the transparency of public policies. The general public can learn whether and how the taxation policy adopted by the government work, thereby helping the government design and improve it. In addition, for researchers, results generated from empirical data-based ex-post evaluations can be conducive to supporting the development and improvement of related theories, especially in verifying the validity of theoretical models \citep{nicolaisen2014ex}. 

Yet, there are three major obstacles to accurately performing ex-post evaluations on the impact of the congestion tax. First, due to \xzhangadd{business proprietary information} and personal privacy, real-world ridesourcing trip data is rarely released to the general public. The paucity of available data prevents researchers from deeply investigating and identifying the effects of such taxes. Second, singling out the precise impact of the congestion tax on transportation system performance (e.g., ridesourcing demand and traffic congestion) is challenging, although not impossible, because transportation system performance is determined by many observed and unobserved factors, such as population and employment changes, road and transit network changes, and periodic changes (seasonality). Identifying the causal link requires sufficient data and high-quality research designs. Third, given the recency of congestion taxes, their implementation generally coincides with the outbreak of the COVID-19 pandemic. Given that studies have shown that COVID-19 has disruptive impacts on ridesourcing, scrutinizing the effects of the congestion tax from the confounding effects of the COVID-19 pandemic would be challenging \citep{yu2022exploring, loa2022has}.

To bridge these gaps, we exploit a natural experiment to empirically examine the short-term impact of congestion taxes on ridesourcing demand and traffic congestion in downtown Chicago. As previously mentioned, Chicago started to enact congestion taxes for ridesourcing on January 6$^{th}$, 2020, but the COVID-19 pandemic did not begin until March 2020. The precious two-month time window between January and March, along with Chicago’s recently released fine-grained ridesourcing trip data and high-frequency traffic speed data, enables us to explore the short-term impact of the policy. To this end, we leverage the ridesourcing trip data to count census tract-level ridesourcing pickups and drop-offs (PUDOs) as the measurement of ridesourcing demand. In addition, we take advantage of traffic segment-level average travel speeds derived from high-frequency traffic speed data to measure traffic congestion. Combined with these data, we employ a regression-based difference-in-differences (DID) causal inference approach as our empirical strategy, in which observations around January 6$^{th}$, 2019 (the year before the implementation of congestion taxes) are adopted as the control group. Lastly, we find that, on average, the congestion tax notably reduces single-trip demand and promotes shared-trip demand. Consequently, the congestion tax significantly curtails overall ridesourcing demand but only has a marginal impact on traffic congestion. These results are consistent and robust against robustness checks. Furthermore, we find that the impact of the congestion tax on ridesourcing demand exhibits heterogeneity across different hours of the day, census tracts, and travel distances.

This study contributes to the literature in the following four ways. First, it offers empirical assessments that add to the ongoing debate on whether and how to regulate ridesourcing using policy instruments. Second, by providing empirical estimations using observational data, this study echoes previous theoretical studies on the effectiveness of per-trip surcharges for ridesourcing, which base their conclusions on abstract mathematical models \citep{li2019regulating,li2021spatial,li2021impact,zhang2022mitigating}. Third, our empirical findings can inform transportation planners and policymakers to frame effective follow-up responses and make optimal policy prescriptions in the future. Lastly, this study provides circumstantial evidence on the relationship between ridesourcing demand and traffic congestion in urban cores, as it demonstrates how traffic congestion will change in response to the reduction of ridesourcing demand.

\xzhangadd{The remainder of the paper is organized as follows. Section \ref{S:2} introduces the congestion tax policy in Chicago. Section \ref{S:3} describes the data sets. The DID model is explicitly explained in Section \ref{S:4}. Section \ref{S:5} presents the results of the DID model, the robustness checks, and the heterogeneity analysis. Section \ref{S:6} summarizes and discusses the major findings, and identifies the follow-up research directions.}

\section{The congestion tax in Chicago}
\label{S:2}
Chicago, as one of the most congested cities in the United States, is severely plagued by traffic congestion problems, particularly in the downtown area. An \textit{INRIX} study shows that Chicago has risen from fifth in 2017 to second in 2019 in the nation in terms of traffic congestion\footnote{\url{https://inrix.com/press-releases/2019-traffic-scorecard-us/}}, with automobile commuters in the city wasting 145 hours a year in congestion. One potential contributor to the increasing traffic congestion is the \xzhangadd{rapid growth} of ridesourcing services. According to a report on ridesourcing and traffic congestion released by the city government of Chicago in October 2019\footnote{\url{https://www.chicago.gov/content/dam/city/depts/mayor/Press\%20Room/Press\%20Releases/2019/October/TNPCongestionReport.pdf}}, the total annual ridesourcing trips (e.g., Uber, Lyft, and Via) in Chicago increased by 271\% from 2015 to 2018. In addition, nearly half of the trips started or ended in the downtown area, and about one-third of these trips both started and ended in the downtown area. This situation results in nearly 26 miles of road space in the downtown area being occupied by ridesourcing vehicles during typical evening rush hours. The report also indicates that the influx of ridesourcing vehicles during peak hours is a crucial factor affecting the travel speed of Chicago Transit Authority (CTA) buses in the downtown area. Accordingly, Chicago Mayor Lori Lightfoot proposed a new congestion tax as an early step in the city’s congestion pricing plan. The congestion tax aims to encourage travelers to share their rides, reduce traffic, save the environment, and promote equity by adjusting the per-trip surcharge for ridesourcing. Thereafter, it was passed by the city council in November 2019 and went into effect on January 6$^{th}$, 2020. The structure of the per-trip surcharge before and after the congestion tax is presented in Table \ref{tab:table1}.

\begin{table}[!ht]
\centering
\caption{The structure of the per-trip surcharge before and after the congestion tax.}
\label{tab:table1}
\resizebox{0.8\textwidth}{!}{%
\begin{tabular}{@{}lllll@{}}
\toprule
Trip type                       & Ride category & Before & After   & Increment \\ \midrule
\begin{tabular}[c]{@{}l@{}}Trips associated with the downtown \\ area in workday peak times \\ (from 6:00 a.m. to 10:00 p.m.)\end{tabular} & Single trip & \$0.72 & \$3 & (+2.28) \\
                                & Shared trip   & \$0.72 & \$1.25  & (+0.53)   \\
                                &               &        & ($-$1.75) &           \\ \midrule
Trips outside the downtown area & Single trip   & \$0.72 & \$1.25  & (+0.53)   \\
                                & Shared trip   & \$0.72 & \$0.65  & ($-$0.07)   \\
                                &               &        & ($-$0.6)  &           \\ \bottomrule
\end{tabular}%
}

\parbox[t]{0.78\textwidth}{\vskip3pt{\footnotesize Notes: Since 2018, for every ridesourcing trip with a pickup or drop-off at O’Hare International Airport, Midway International Airport, Navy Pier, or McCormick Place, there is an additional \$5 charge. In addition, the per-trip surcharge for all wheelchair accessible vehicle (WAV) trips decreased from \$0.62 to \$0.55 after the congestion tax. Given that WAV trips only account for a negligible fraction of all ridesourcing trips, we do not specifically consider these trips.}}

\end{table}

As Table \ref{tab:table1} shows, before the implementation of the congestion tax, the per-trip surcharge has a flat structure. At any time, it is \$0.72 for all trips, regardless of whether the origin or destination is in the downtown area and whether the trip is shared. After the implementation of this tax, the per-trip surcharge changed to a diverse structure. For trips associated with the downtown area (i.e., those starting or ending within the downtown area shown in Fig. \ref{fig:fig1}) in workday peak times (6:00 a.m. – 10:00 p.m.), the per-trip surcharge increased from \$0.72 to \$3.00 for single trips but only increased to \$1.25 for shared trips. Although the surcharge for single and shared trips are adjusted upward by the congestion tax, the magnitudes of the adjustments are not at the same level. Shared trips were taxed \$1.75 less than single trips. For trips outside the downtown, the surcharge increased from \$0.72 to \$1.25 for single trips but decreased to \$0.65 for shared trips. 

Compared with the ridesourcing tax in other cities, Chicago's congestion tax is characterized by (1) the per-trip instead of \textit{ad valorem} surcharge, (2) the extra fee on the trips associated with the downtown area, and (3) the discount for shared trips (interested readers can refer to \cite{lehe2021taxation}, which provides a comprehensive profile of existing ridesourcing tax schemes in the United States). Given that the downtown area is the primary focus of the congestion tax and attracts nearly half of all ridesourcing trips, we exclusively perform an analysis on the impact of the congestion tax on ridesourcing demand and traffic congestion within the downtown area in this study.

\section{Data}
\label{S:3}

We use two primary data sets obtained from the Chicago data portal\footnote{\url{https://data.cityofchicago.org/}} to measure ridesourcing demand and traffic congestion, along with weather data collected from the Weather Underground website\footnote{\url{https://www.wunderground.com/history/daily/us/il/chicago}} as control variables.

\subsection{Ridesourcing trip data}
Chicago has recently opened its ridesourcing trip data set to the general public. The data set contains all ridesourcing trips that occurred in Chicago starting from November 2018. In particular, it records each trip’s start and end time, the geographical location of the origin and destination, whether travelers are willing to share the trip, and whether sharing-authorized trips are matched, etc. To protect user privacy, the start time is rounded to the nearest 15 minutes, and the start and end locations are aggregated at the census tract level. In addition, if there are only two or fewer trips in a census tract within a 15-minute time window, the start and end locations of these trips are aggregated at the community area level (a larger geographic unit than the census tract) to prevent them from being identiﬁed. As a result, about one-third of the trips have census tract information hidden and arbitrarily removing these trips may lead to biased model results. To tackle this problem, we employ the stratified sampling procedure proposed by \cite{xu2021identifying} to infer the hidden census tracts. A total of 29 census tracts within the downtown area are selected as the basic spatial unit of the subsequent analysis, as shown in Fig. \ref{fig:fig2}. For these census tracts, we count the hourly census tract-level PUDOs of various types of ridesourcing trips (i.e., single, sharing-authorized, sharing-matched, and all trips) to comprehensively measure ridesourcing demand, which is informed by previous studies \citep{zhang2022machine, yu2019exploring, dean2021spatial}.

\subsection{Traffic speed data}

\begin{figure}[!ht]
     \centering
     \begin{subfigure}[b]{0.425\textwidth}
         \centering
         \includegraphics[width=\textwidth]{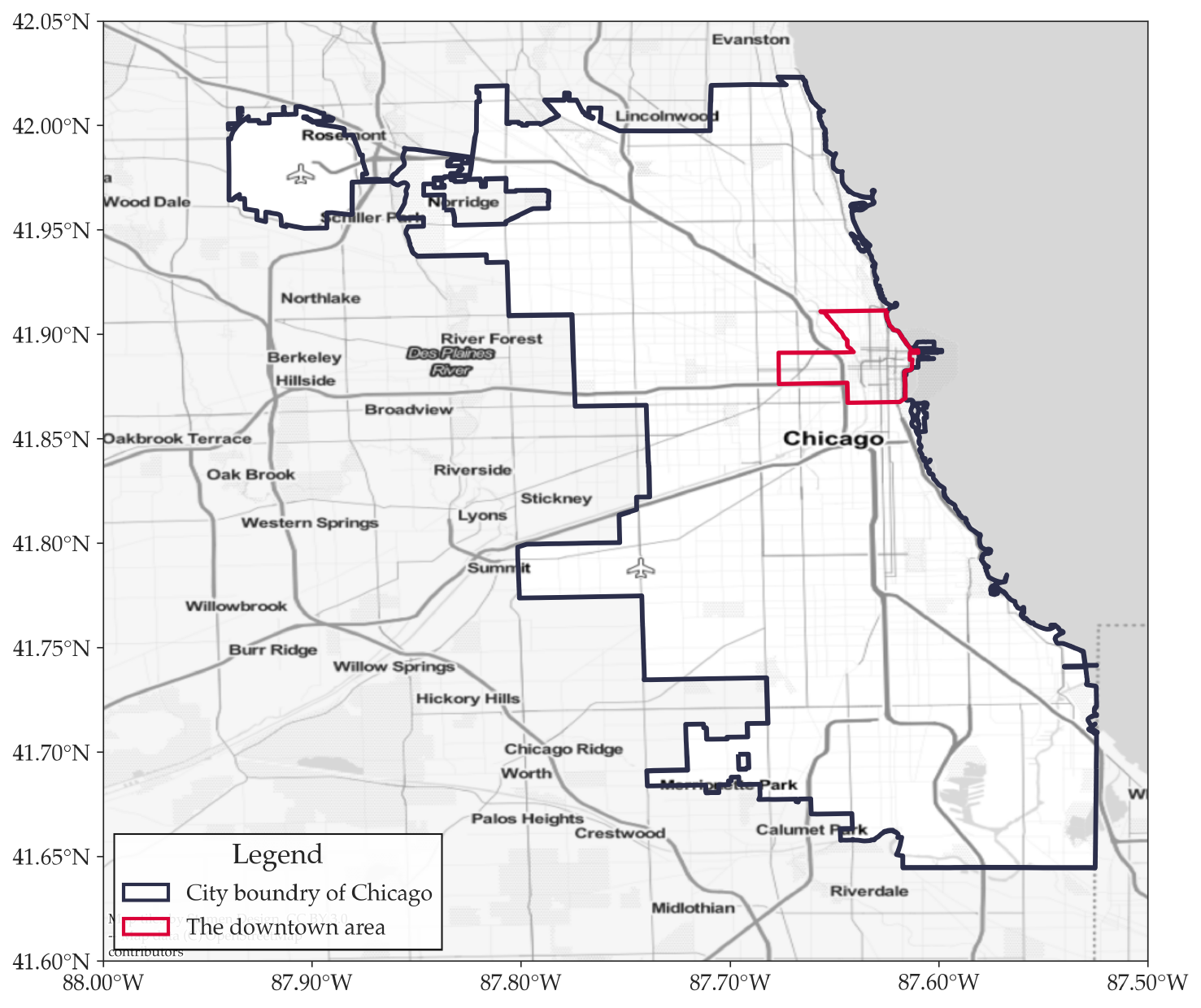}
         \caption{The downtown area of Chicago}
         \label{fig:fig1}
     \end{subfigure}
     \hfill
     \begin{subfigure}[b]{0.5\textwidth}
         \centering
         \includegraphics[width=\textwidth]{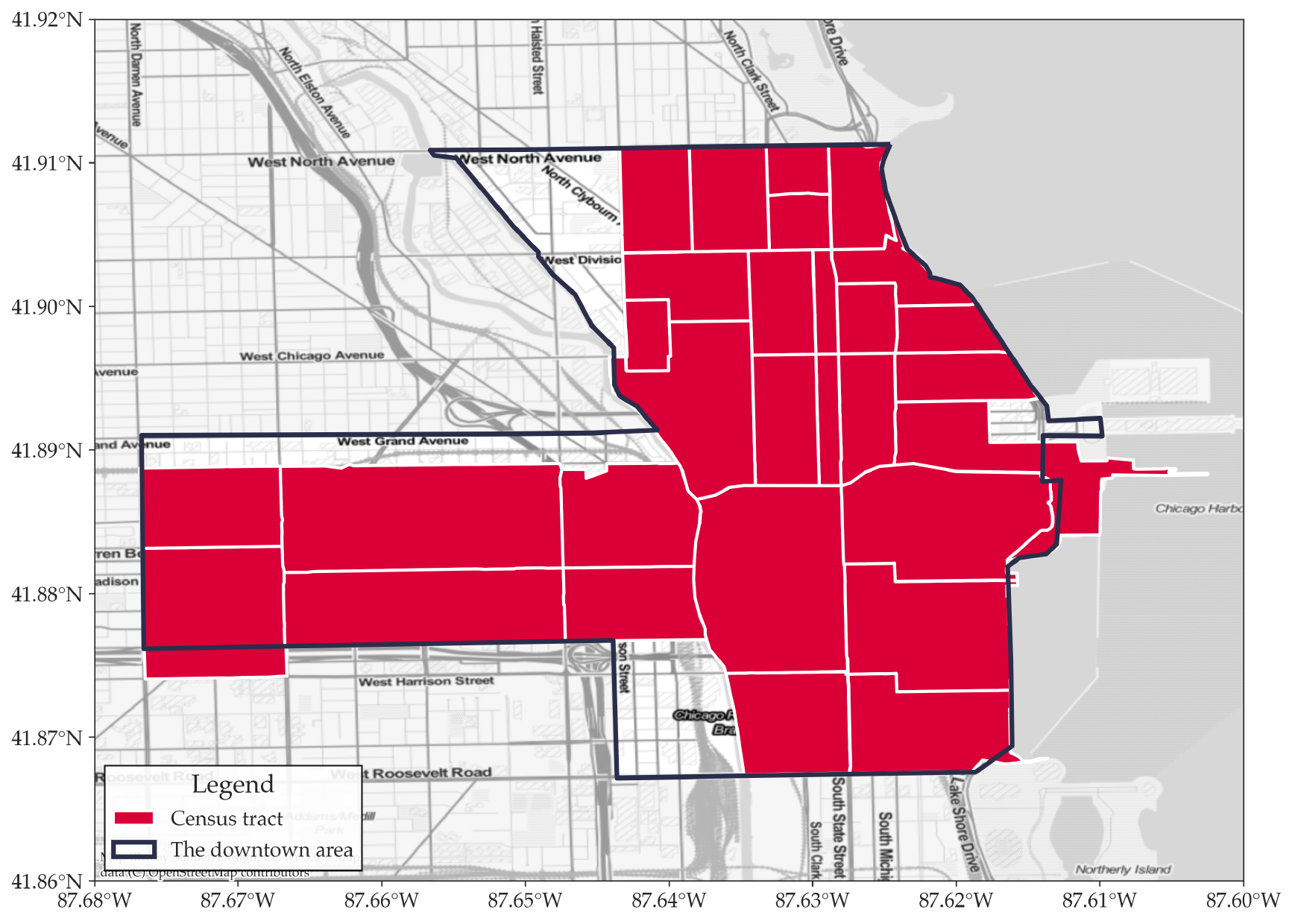}
         \caption{Selected census tracts within the downtown area.}
         \label{fig:fig2}
        \end{subfigure}
    
    \begin{subfigure}[b]{0.475\textwidth}
    \centering
    \includegraphics[width=\textwidth]{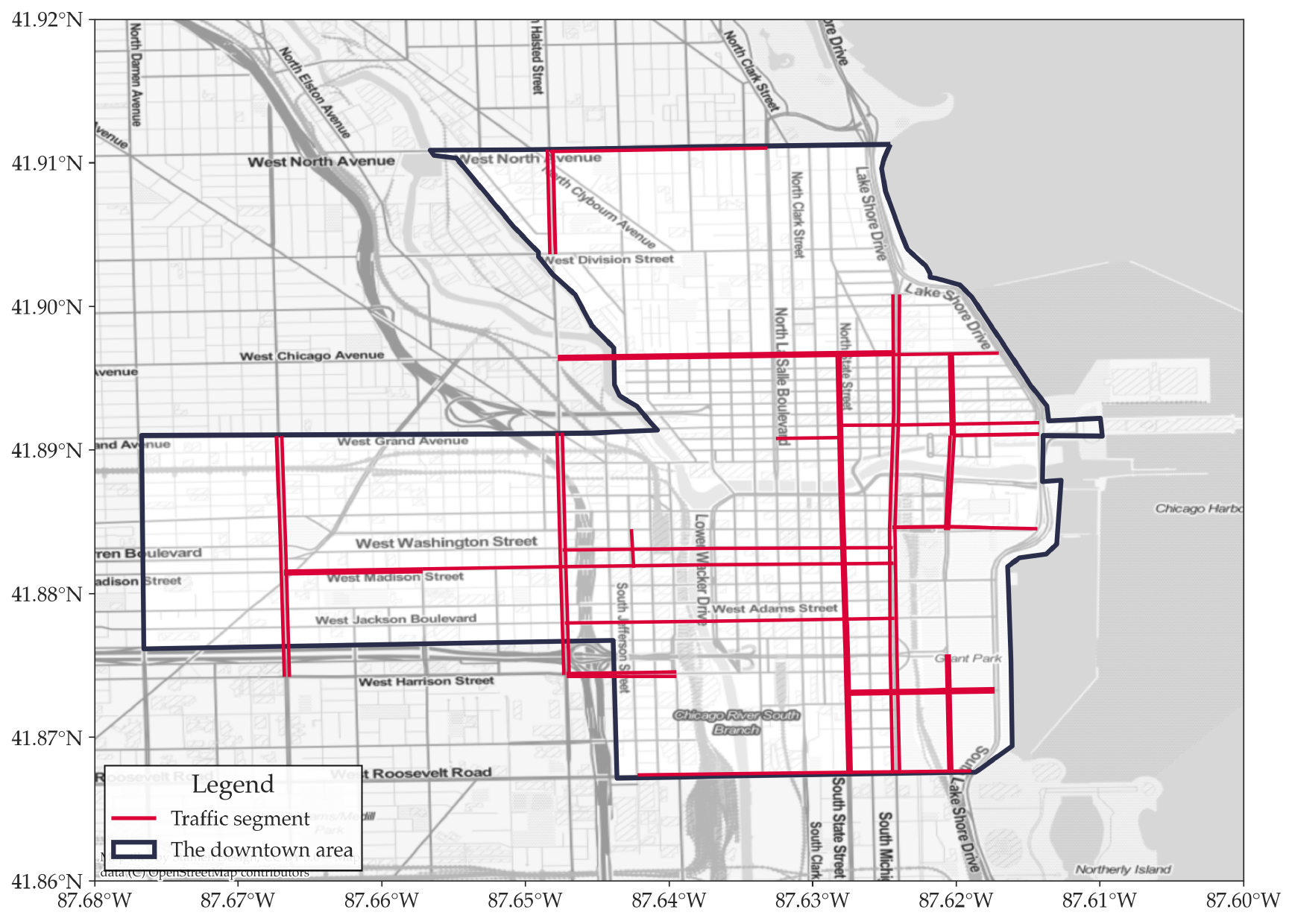}
    \caption{Selected traffic segments within the downtown area}
    \label{fig:fig3}
     \end{subfigure}
     
    \caption{Studied areas: census tracts and traffic segments}
    \label{fig:fig123}
\end{figure}

The traffic speed data set provides the historical traffic speed in miles per hour of approximately 130 traffic segments within the downtown area of Chicago from 2018 to date. These traffic segments are almost arterial streets (non-highway streets) and typically one-half mile long. The traffic speed is estimated based on GPS trajectories received from Chicago Transit Authority (CTA) buses, which are updated every five minutes. However, the estimated traffic speed depends on the number of CTA buses on the street. If the number of CTA buses on the street is sparse, no speed will be recorded. Therefore, we exclude the segments where traffic speeds are not consistently monitored. Hence, a total of 59 traffic segments within the downtown area were retained for the subsequent analysis, as shown in Fig. \ref{fig:fig3}. Lastly, we aggregate traffic speeds at the traffic-segment-by-hour level and take the mean value as the measurement of traffic congestion.

\subsection{Weather data}
\label{S:3.2}

We collect the hourly weather data of Chicago from the Weather Underground website\footnote{\url{https://www.wunderground.com/history/daily/us/il/chicago}} as control variables for the subsequent analysis. The weather data contains the hourly temperature and precipitation/snowfall level. In particular, temperature is a continuous variable with the unit of Fahrenheit. Precipitation/snowfall level is assessed using a set of dummy variables with seven categories: no rain/snow, light drizzle, light rain, rain, heavy rain, light snow, and snow.

\subsection{Data cleaning strategy}

Given that the congestion tax only takes effect during workday peak times, we first exclude weekend and holiday samples\footnote{Holidays include Veteran’s Day, Thanksgiving Day, Day After Thanksgiving Day, Days between December $25^{th}$ and January $3^{rd}$ (Christmas Day and New Year’s Day), Birthday of Martin Luther King, Jr., Lincoln’s Birthday, and Washington’s Birthday.}. Days with unusual ridesourcing demand patterns are also excluded (e.g., the day before  Thanksgiving Day and Christmas Day, Valentine’s Day). Moreover, we remove the samples in the first week after implementing the congestion tax. The reason is that in the first week after its implementation, some travelers \xzhangadd{may be} still unaware of the congestion tax and thus hardly make an immediate response. Including samples in this period will lead to biased model results (\xzhangadd{we formally test this assertion in the section of \ref{appendix_a}}). 

\section{DID: a causal inference approach}
\label{S:4}

The most intuitive approach to examining the short-term impact of the congestion tax on ridesourcing demand and traffic congestion is to conduct before-after comparisons. However, ridesourcing demand and traffic congestion often have evident seasonal changes driven by unobserved factors. Simple before–after comparisons likely confound the causal impact of the congestion tax with seasonal changes. To perform reliable causal inference, we employ the DID modeling approach. The basic idea of the DID modeling approach is to mimic an experimental research design by leveraging panel data. In DID models, samples are divided into treatment and control groups. The underlying assumption is that the treatment group is directly exposed to the exogenous shock (e.g., the enforcement of a policy). By contrast, the control group is unaffected by the exogenous shock and thus provides a counterfactual scenario. By estimating the differences between the treatment and control groups in the changes in the outcome variable before and after the exogenous shock, the causal impact of the exogenous shock on the outcome variable can be inferred. 

It is worth noting that as a causal inference framework, the DID model setting is highly flexible. As aforementioned, in its conventional form, analysis units (census tracts in this study) are divided into treatment and control groups based on whether they are exposed to an exogenous shock. However, if the exogenous shock is \xzhangadd{universal} and \xzhangadd{does not target specific groups} or areas, there would be no eligible control groups. Taking the congestion tax in Chicago as an example, not only the per-trip surcharges on ridesourcing trips associated with the downtown area but also those outside the downtown area are raised after the implementation of the policy (although in varying degrees). This feature means that ridesourcing demand is affected for the entire city, not just the downtown area. Accordingly, the conventional DID model setting is not applicable because we cannot find “unpolluted” areas as control groups. To tackle this challenge, a \xzhangadd{smarter and subtler} form of DID models is introduced in this study. To be specific, we take samples within the \xzhangadd{60-day [-30 days, +30 days]} symmetrical time window\footnote{The reason why we choose the \xzhangadd{60-day} symmetrical time window is to make a trade-off between the sample size and interference of COVID-19, given that ridesourcing demand started to decrease in March 2020 because of the COVID-19 pandemic.} before and after January 6$^{th}$, 2020 (the day that the congestion tax goes into effect) as the treatment group. At the same time, samples within the \xzhangadd{60-day} symmetrical time window before and after January 6$^{th}$, 2019, are adopted as the control group. In other words, we believe that samples around January 2019 could provide a counterfactual scenario of what 2020 ridesourcing demand and traffic congestion would be if there were no congestion taxes, due to the following reasons. First, the travel pattern during the same season in different years usually has similar trends. Second, there have been only trivial changes (from \$0.67 to \$0.72) in the per-trip surcharge for ridesourcing trips since January 2019, which are largely not enough to impact ridesourcing demand significantly. Moreover, a number of rigorous studies published in top journals have employed such DID model setting (using last year’s samples as the control group), soundly justifying the validity of this approach \citep{tarduno2021congestion,chen2022jue,gibson2015effects,he2020short}.

Mathematically, our DID model can be written as follows:
\begin{equation}
\label{eq1}
\log \left(y_{i, t, h}\right)=\beta_{0}+\beta_{1} \operatorname{POST}_{t}+\beta_{2} T R E A T_{t}+\beta_{3} \text { POST }_{t} \times T R E A T_{t}+\beta_{4} Z_{t, h}+\delta_{i}+\varepsilon_{i, t, h}
\end{equation}

\begin{figure}[!h]
    \centering
    \includegraphics[width=0.8\textwidth]{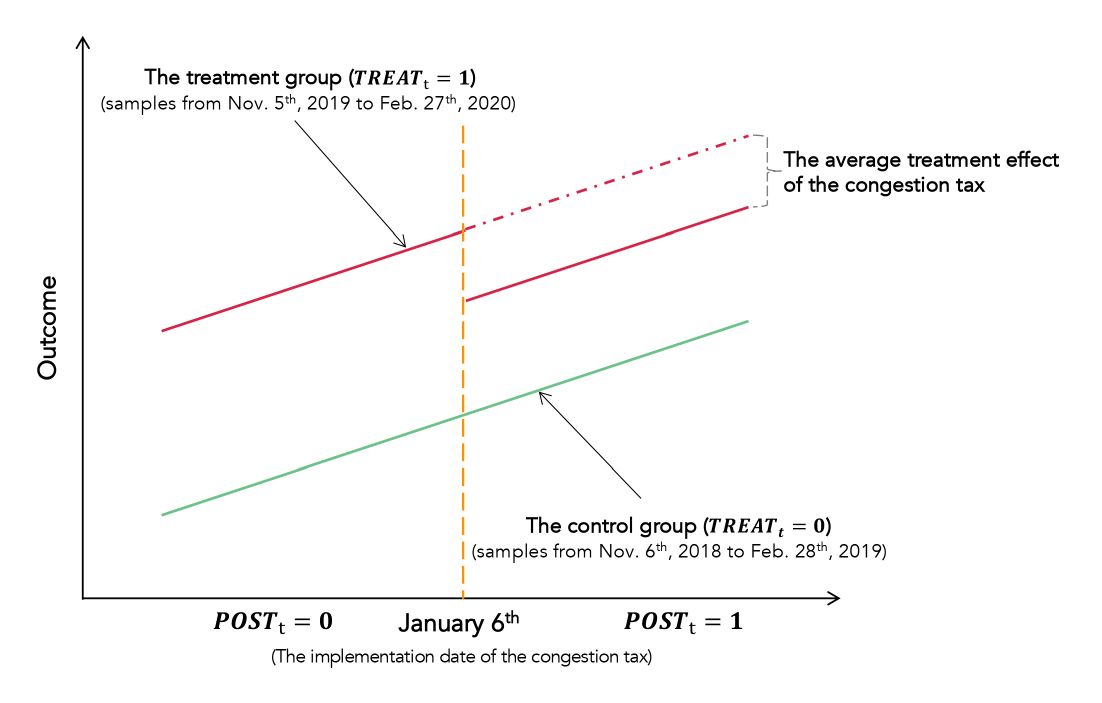}
    \caption{\xzhangadd{The graphical illustration of the DID model setting used in this study.}}
    \label{fig:fig4}
\end{figure}

\noindent Where $y_{i,t,h}$ is the outcome variable (i.e., PUDOs and traffic speed) of census tract or traffic segment $i$ on hour $h$ of day $t$. $POST_t$ is a dummy variable with the value of one if day $t$ falls in the time period between January $6^{th}$ and February $27^{th}$/$28^{th}$, 2019 or 2020, and zero otherwise. To put it more simply, $POST_t$ represents whether the policy is implemented. $TREAT_t$ is an indicator variable for the treatment group. It equals one if day $t$ falls in the time period between November $5^{th}$, 2019 and February $28^{th}$, 2020, and zero otherwise. $\beta_3$, the coefficient of the interaction term of $POST_t$ and $TREAT_t$, is the parameter of our primary interest. We can calculate the short-term percentage point impact of the congestion tax on ridesourcing demand or traffic congestion as $\% \Delta y=e^{\beta_{3}}-1$. $Z_{t,h}$ is the vector of control variables, including the temperature, precipitation/snowfall level, hour of day, and day of week. $\delta_i$ is the census tract or traffic segment fixed effects. $\varepsilon_{i,t,h}$ denotes the error term. To account for potential within-group error correlations, we cluster the standard errors at the census-tract-by-day or traffic-segment-by-day level. For ease of understanding, Fig. \ref{fig:fig4} offers the graphical illustration of Eq. \ref{eq1}. And Table \ref{tab: Summary statistics} presents the summary statistics of the samples.

\newpage

\begin{table}[!h]
\centering
\caption{Summary statistics.}
\label{tab: Summary statistics}
\resizebox{0.8\textwidth}{!}{%
\begin{tabular}{llllll}
\hline
                                 & \multirow{2}{*}{All} & \multicolumn{2}{l}{Treatment group} & \multicolumn{2}{l}{Control group} \\ \cline{3-6} 
                             &          & Before   & After    & Before   & After    \\ \hline
\textit{Outcome variables}   &          &          &          &          &          \\
PUDOs of all trips                & 343.54               & 365.32           & 324.23           & 344.46          & 340.14          \\
                             & (404.21) & (419.87) & (385.70) & (400.36) & (409.14) \\
PUDOs of single trips             & 292.19               & 335.10           & 285.32           & 275.73          & 272.61          \\
                             & (347.28) & (386.99) & (342.68) & (321.90) & (330.27) \\
PUDOs of sharing-authorized trips & 51.35                & 30.21            & 38.91            & 68.73           & 67.53           \\
                             & (67.58)  & (34.62)  & (44.82)  & (82.98)  & (83.74)  \\
PUDOs of sharing-matched trips   & 40.17                & 22.61            & 28.63            & 53.63           & 55.81           \\
                             & (57.36)  & (27.98)  & (35.68)  & (69.46)  & (73.58)  \\
Traffic speed                & 19.51    & 19.52    & 19.72    & 19.38    & 19.43    \\
                             & (3.49)   & (3.53)   & (3.50)   & (3.46)   & (3.45)   \\
\textit{Control variables}   &          &          &          &          &          \\
Temperature                  & 30.29    & 34.06    & 30.98    & 33.33    & 22.79    \\
                             & (10.58)  & (9.72)   & (6.39)   & (6.81)   & (13.63)  \\
Precipitation/snowfall level &          &          &          &          &          \\
No rain/snow                 & 82.1\%   & 92.9\%   & 75.0\%   & 82.7\%   & 77.9\%   \\
Light drizzle                & 1.7\%    & 2.3\%    & 0.8\%    & 1.1\%    & 2.5\%    \\
Light rain                   & 2.0\%    & 2.0\%    & 1.7\%    & 1.9\%    & 2.3\%    \\
Rain                         & 0.1\%    & 0.3\%    & 0\%      & 0.2\%    & 0\%      \\
Heavy rain                   & 0.1\%    & 0.4\%    & 0\%      & 0\%      & 0\%      \\
Light snow                   & 12.4\%   & 2.1\%    & 21.3\%   & 13.5\%   & 12.9\%   \\
Snow                         & 1.6\%    & 0.0\%    & 1.2\%    & 0.6\%    & 4.4\%    \\ \hline
\end{tabular}%
}
\parbox[t]{0.78\textwidth}{\vskip3pt{\footnotesize Notes: Each Column summarizes the mean values and standard deviations (in parenthesis) of different variables at the hourly level during workday peak times (from 6:00 a.m. to 10:00 p.m.).}}

\end{table}

One crucial assumption for the DID model is that the outcome variables of the treatment and control groups are supposed to have parallel trends in the absence of the exogenous shock (the congestion tax in this context). Otherwise, the control group cannot provide a reliable counterfactual reference, and thus the impact of the exogenous shock will be misestimated. To formally test whether the parallel trend assumption is valid in this study, we divide the \xzhangadd{60-day} symmetrical time window before and after January $6^{th}$ into twelve 5-day intervals for both treatment and control groups and run the following model:
\begin{equation}
\label{eq2}
\log \left(y_{i, t, h}\right)=\beta_{0}+\beta_{1} \text { POST }_{t}+\beta_{2} T R E A T_{t}+\sum_{k \neq-1} \beta_{3 k} I_{k t} T R E A T_{t}+\beta_{4} Z_{t, h}+\delta_{i}+\varepsilon_{i, t, h}
\end{equation}

\noindent Where $I_{kt}$ equals one if day t falls in the $k$-th 5-day interval around January $6^{th}$, 2019 or 2020, and zero otherwise. The first 5-day interval before January $6^{th}$, 2019 or 2020, is adopted as the reference ($k\neq-1$). Other terms are the same as in Eq. \ref{eq1}. $\beta_{3k}$ are the coefficients of interest. 

\begin{figure}[!ht]
     \centering
     \begin{subfigure}[b]{0.475\textwidth}
         \centering
         \includegraphics[width=\textwidth]{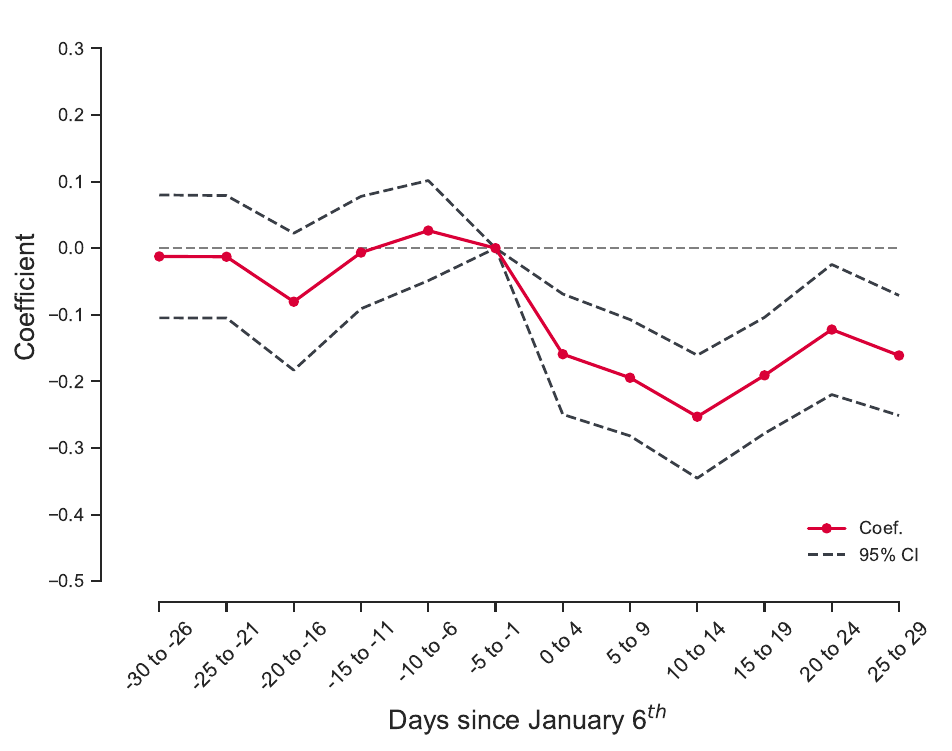}
         \caption{PUDOs of single trips}
         \label{fig:PUDO of single trips}
     \end{subfigure}
     \hfill
     \begin{subfigure}[b]{0.475\textwidth}
         \centering
         \includegraphics[width=\textwidth]{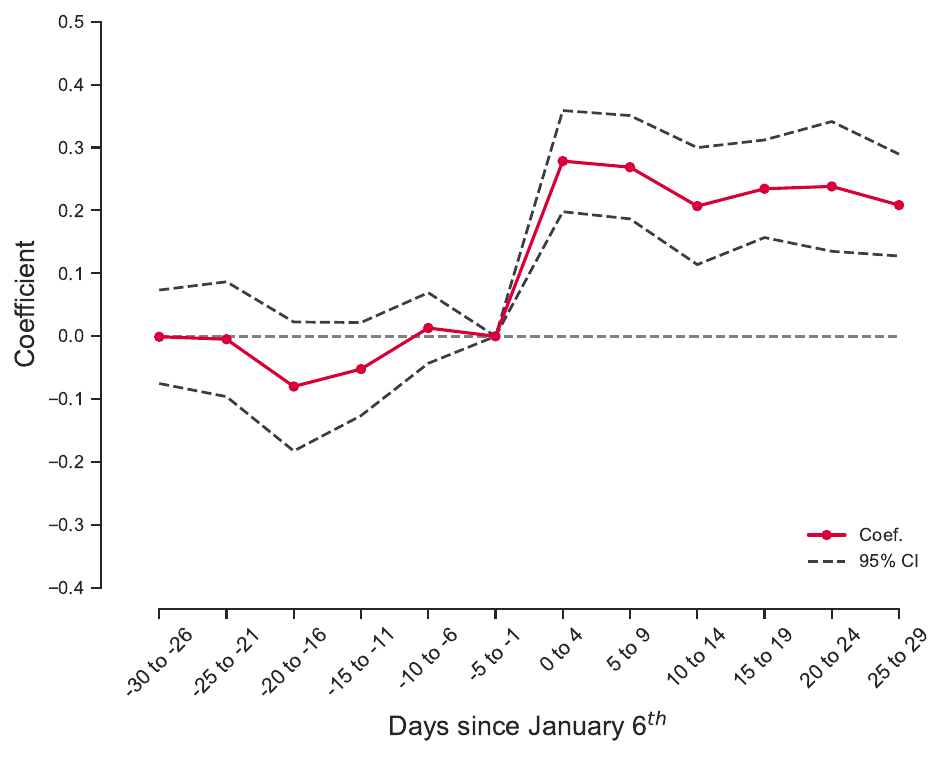}
         \caption{PUDOs of sharing-authorized trips}
         \label{fig:PUDO of sharing-authorized trips}
        \end{subfigure}
    
    \begin{subfigure}[b]{0.475\textwidth}
    \centering
    \includegraphics[width=\textwidth]{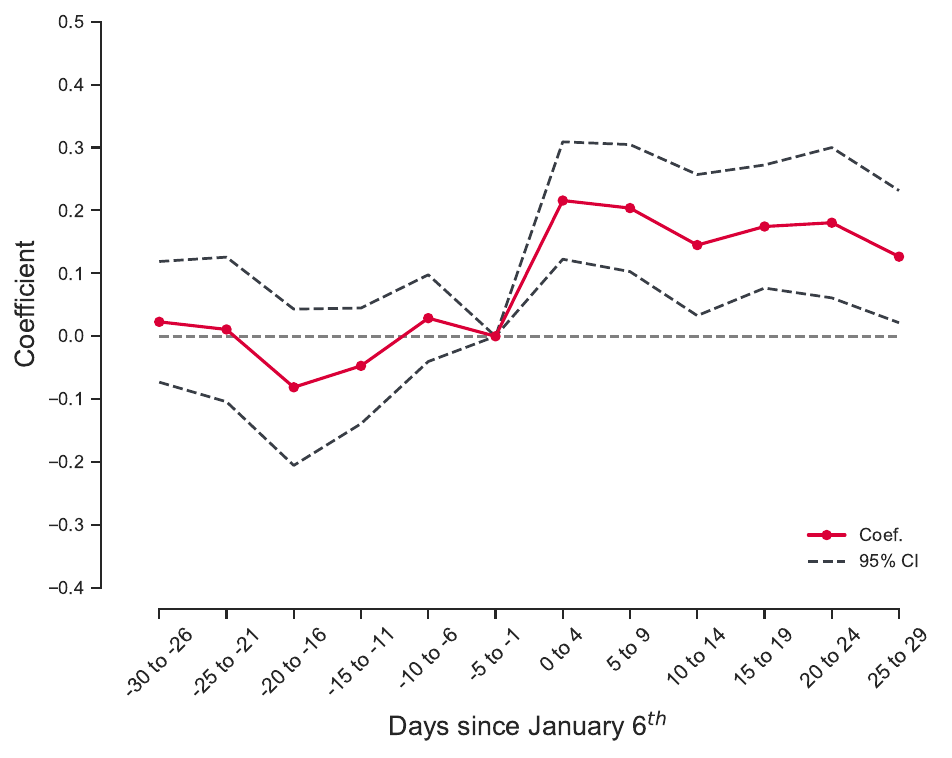}
    \caption{PUDOs of sharing-matched trips}
    \label{fig:PUDO of sharing-matched trips}
     \end{subfigure}
     \hfill
     \begin{subfigure}[b]{0.475\textwidth}
    \centering
    \includegraphics[width=\textwidth]{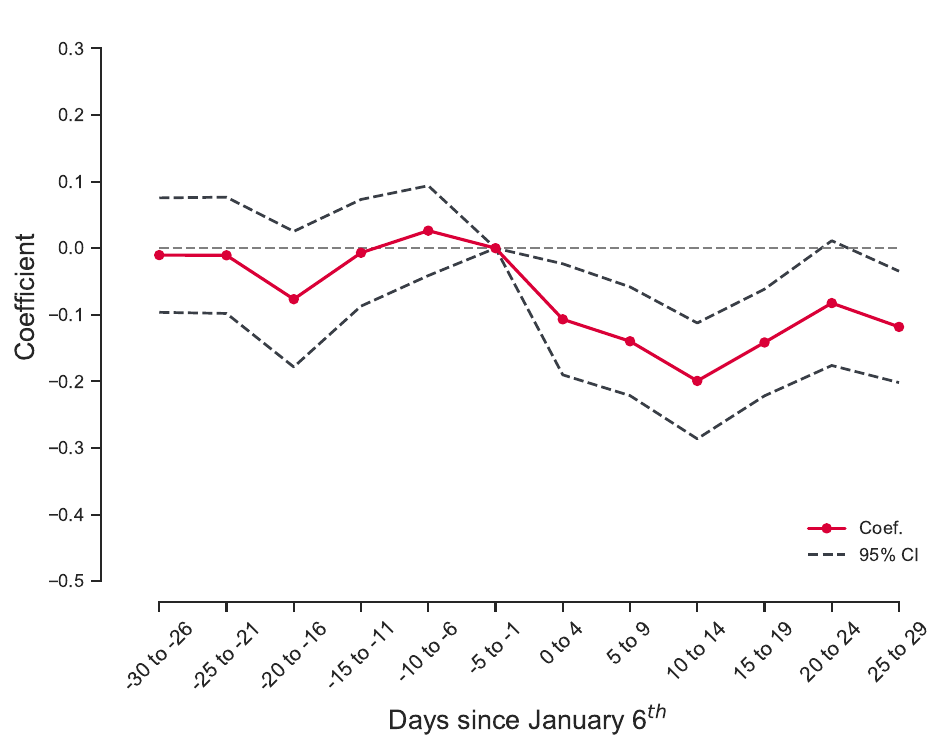}
    \caption{PUDOs of all trips}
    \label{fig:PUDO of all trips}
     \end{subfigure}

    \begin{subfigure}[b]{0.475\textwidth}
    \centering
    \includegraphics[width=\textwidth]{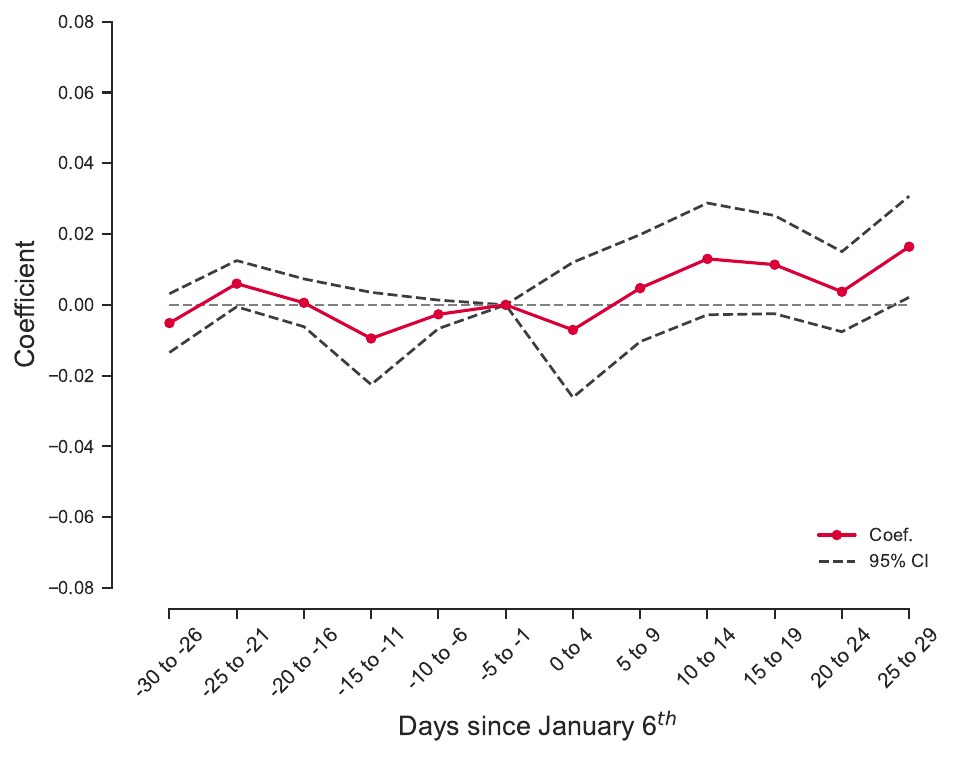}
    \caption{Traffic speeds}
    \label{fig:Traffic speeds}
     \end{subfigure}
     
    \caption{Results of parallel trend assumption test}
    \label{fig:fig5}
\end{figure}

Fig. \ref{fig:fig5} presents the estimated coefficients of $\beta_{3k}$ for five outcome variables: PUDOs of all trips, PUDOs of single trips, PUDOs of sharing-authorized trips, PUDOs of sharing-matched trips, and traffic speeds. We observe that among all five models, for 5-day intervals before January $6^{th}$ ($k<0$), their coefficients are all insignificant (confidence intervals straddle zero). In other words, compared to the reference interval, the changes in the outcome variables between the treatment and control groups have no significant differences for all intervals before the implementation of the congestion tax. This observation suggests that the parallel trend assumption for DID models is reasonable.

\section{Results}
\label{S:5}

\subsection{Baseline results}
\label{S:5.1}
Table \ref{tab:Baseline results} reports our baseline results from Eq. \ref{eq1}, pooling across hours of workday peak times (from 6:00 a.m. to 10:00 p.m.). The estimated coefficient of TREAT×POST captures the average impact of the congestion tax on the outcome variable. We observe that PUDOs of single trips decrease by 14.7\% ($e^{-0.159}-1$, in Column 2), but PUDOs of sharing-authorized trips increase by 30.3\% ($e^{0.265} - 1$, in Column 3) following the implementation of the congestion tax. This result suggests that there may be some shifting of single trips towards sharing-authorized trips. In addition, there is a 21.2\% increase ($e^{0.192} - 1$, in Column 4) in PUDOs of sharing-matched trips. These outcomes imply that the congestion tax significantly reduces single-trip demand and raises travelers’ willingness to share their trips. As more travelers would like to share their trips, successfully matched trips also increased. Consequently, we find that, on average, the congestion tax significantly reduces PUDOs of all ridesourcing trips by 10.6\% ($1-e^{-0.112}$, in Column 1). At the same time, Column 5 shows that although with a positive sign, the impact of the congestion tax on traffic speeds is of a relatively small magnitude (only 0.008) and statistically insignificant. This result indicates that the congestion tax considerably curbs overall ridesourcing demand but has a null impact on reducing traffic congestion. In summary, our baseline results offer evidence that, in the short term, the congestion tax achieves its goal of reducing ridesourcing demand and promoting shared trips within the downtown area. However, the positive impact of the congestion tax on traffic speeds is marginal and statically insignificant, indicating its trivial role in alleviating traffic congestion.

\begin{table}[!ht]
\centering
\caption{Baseline results.}
\label{tab:Baseline results}
\resizebox{\textwidth}{!}{%
\begin{tabular}{@{}llllll@{}}
\toprule
                              & \multicolumn{4}{l}{Log(PUDOs+1)}                                             & \multirow{2}{*}{Log(traffic speeds)} \\ \cmidrule(lr){2-5}
                              & All trips & Single trips & Sharing-authorized trips & Sharing-matched trips &                                      \\ \cmidrule(l){2-6} 
                              & (1)       & (2)          & (3)                      & (4)                   & (5)                                  \\ \midrule
TREAT×POST                    & -0.112*** & -0.159***    & 0.265***                 & 0.192***              & 0.008                                \\
                              & (0.020)   & (0.022)      & (0.026)                  & (0.033)               & (0.007)                              \\
TREAT                         & 0.043**   & 0.186***     & -0.774***                & -0.800***             & 0.005                                \\
                              & (0.018)   & (0.019)      & (0.020)                  & (0.028)               & (0.006)                              \\
POST                          & -0.022    & -0.017       & -0.045**                 & 0.008                 & 0.007                                \\
                              & (0.019)   & (0.020)      & (0.020)                  & (0.026)               & (0.006)                              \\
Weather control               & Y         & Y            & Y                        & Y                     & Y                                    \\
Hour of day fixed effects     & Y         & Y            & Y                        & Y                     & Y                                    \\
Day of week fixed effects     & Y         & Y            & Y                        & Y                     & Y                                    \\
Census tract fixed effects    & Y         & Y            & Y                        & Y                     &                                      \\
Traffic segment fixed effects &           &              &                          &                       & Y                                    \\
No. of observations           & 55,680    & 55,680       & 55,680                   & 55,680                & 113,280                              \\
Adj. R-squared                & 0.938     & 0.935        & 0.909                    & 0.894                 & 0.619                                \\ \bottomrule
\end{tabular}%
}
\parbox[t]{0.98\textwidth}{\vskip3pt{\footnotesize Notes: Standard errors clustered at the census-tract-by-day (or traffic-segment-by-day) level are provided in parenthesis. *Significant at the 10\% level, **Significant at the 5\% level, ***Significant at the 1\% level.}}
\end{table}

\begin{figure}[!ht]
     \centering
     \begin{subfigure}[b]{0.475\textwidth}
         \centering
         \includegraphics[width=\textwidth]{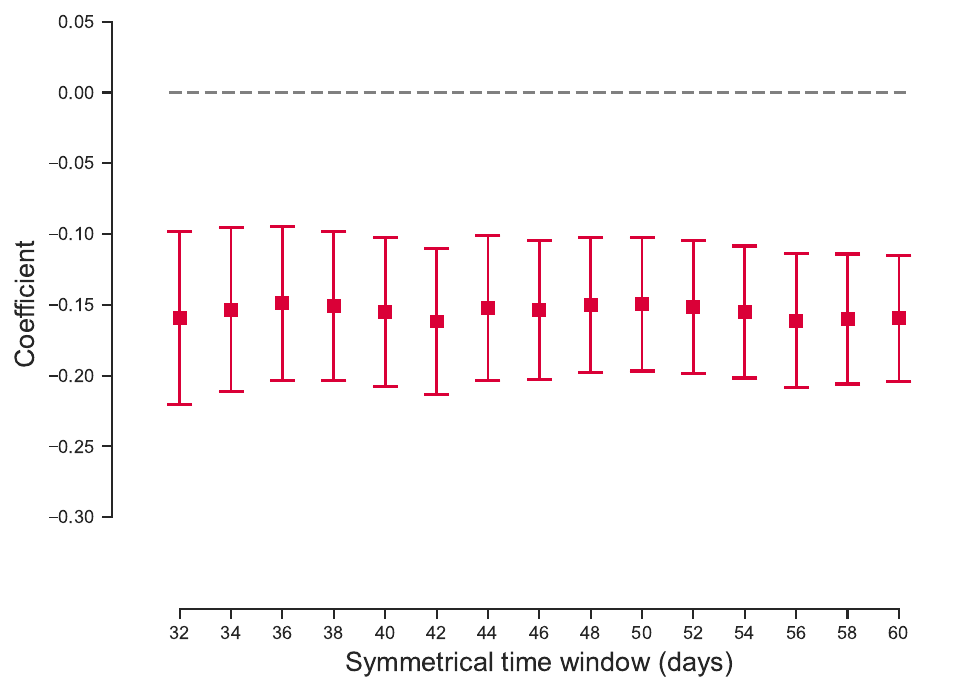}
         \caption{PUDOs of single trips}
         \label{fig:tw_PUDO of single trips}
     \end{subfigure}
     \hfill
     \begin{subfigure}[b]{0.475\textwidth}
         \centering
         \includegraphics[width=\textwidth]{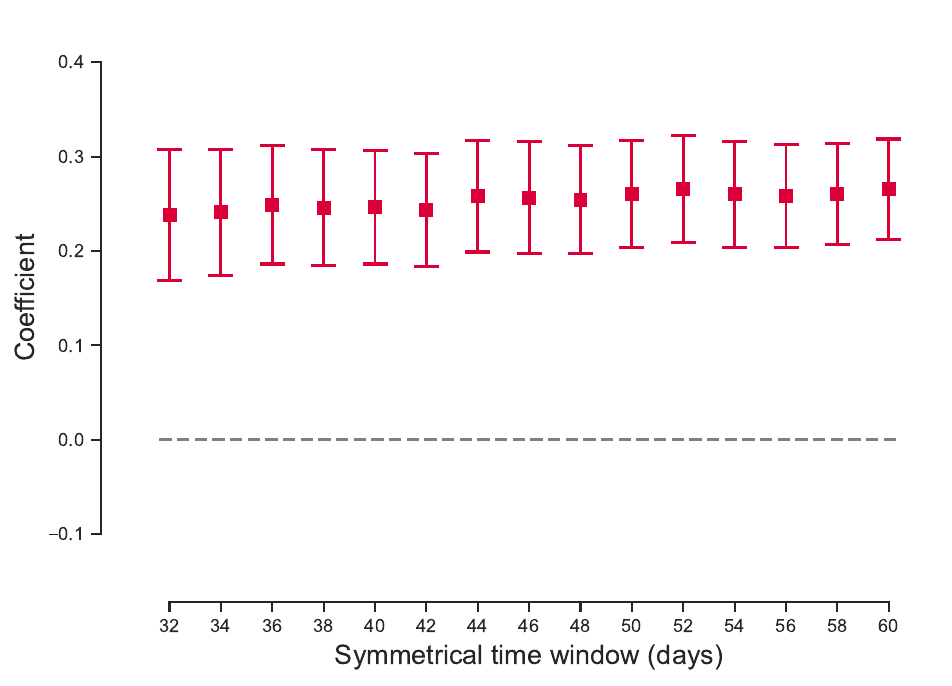}
         \caption{PUDOs of sharing-authorized trips}
         \label{fig:tw_PUDO of sharing-authorized trips}
        \end{subfigure}
    
    \begin{subfigure}[b]{0.475\textwidth}
    \centering
    \includegraphics[width=\textwidth]{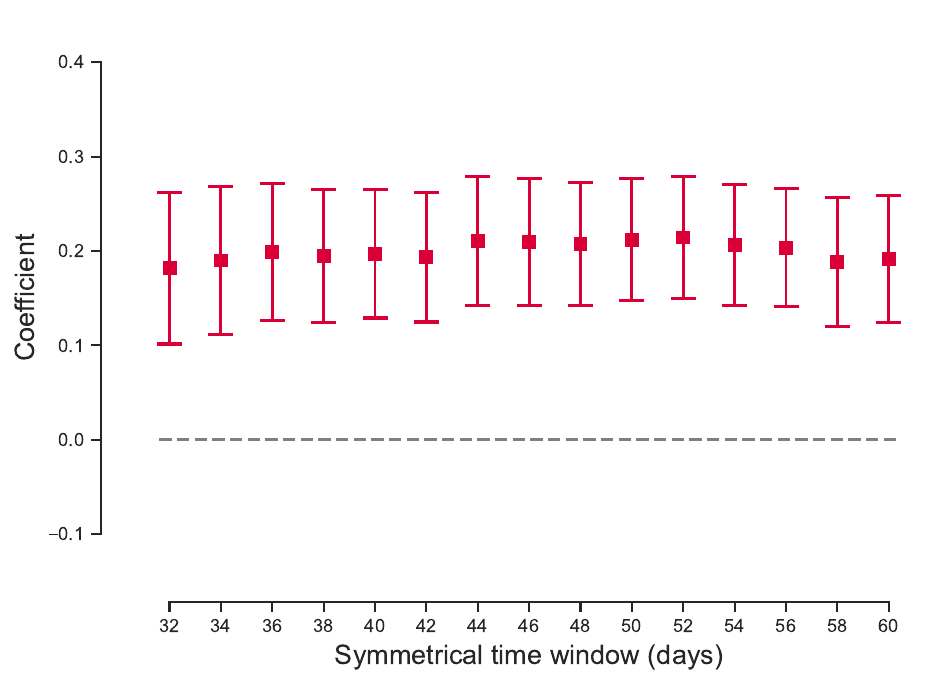}
    \caption{PUDOs of sharing-matched trips}
    \label{fig:tw_PUDO of sharing-matched trips}
     \end{subfigure}
     \hfill
     \begin{subfigure}[b]{0.475\textwidth}
    \centering
    \includegraphics[width=\textwidth]{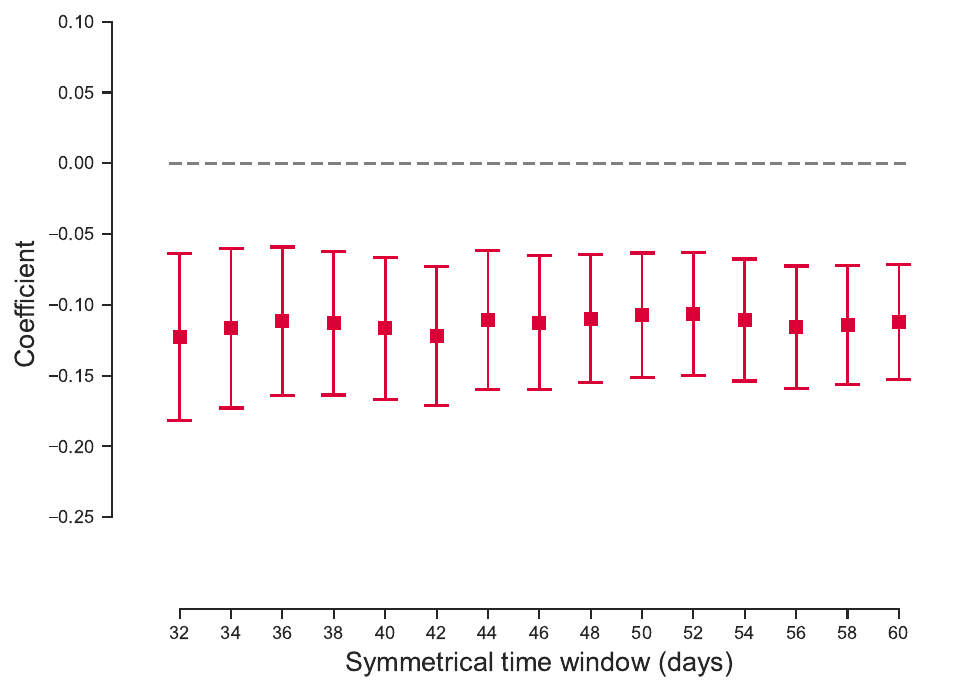}
    \caption{PUDOs of all trips}
    \label{fig:tw_PUDO of all trips}
     \end{subfigure}

    \begin{subfigure}[b]{0.475\textwidth}
    \centering
    \includegraphics[width=\textwidth]{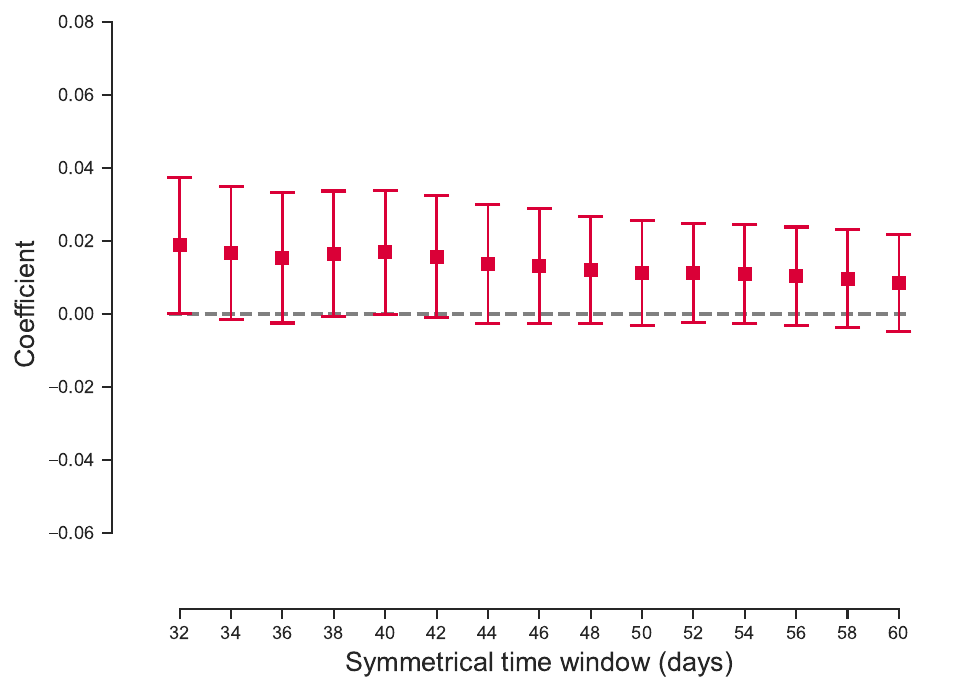}
    \caption{Traffic speeds}
    \label{fig:tw_Traffic speeds}
     \end{subfigure}
     
    \caption{Results of the time window sensitivity test}
    \label{fig:Results of the time window sensitivity test}
\end{figure}

\subsection{Robustness checks}

In Section \ref{S:5.1}, we have identified the impact of the congestion tax on ridesourcing demand and traffic congestion using the DID modeling approach. To ensure that our identifications are robust and reliable, we perform several robustness checks. 

First, we conduct a time window sensitivity test to show that our baseline results are stable over different time windows. Fig. \ref{fig:Results of the time window sensitivity test} plots the estimated coefficients of interest and their 95\% confidence intervals from Eq. \ref{eq1} for symmetrical time windows ranging from \xzhangadd{32} to \xzhangadd{60} days before and after January $6^{th}$. We can see that the estimated coefficients are significantly negative across all time windows for PUDOs of single and all trips. By contrast, the estimated coefficients of PUDOs of sharing-authorized and sharing-matched trips are significantly positive. For traffic speeds, nearly all estimated coefficients are insignificant (the confidence intervals span zero) across all time windows. Moreover, we find that the magnitude of the estimated coefficients for all five outcome variables only has a slight variation as the time window varies. The above results show that our baseline results are robust against the choice of time windows.

Second, we expand our analysis to include all traffic segments to demonstrate that our baseline results for traffic speeds are not driven by our data cleaning strategy. As mentioned in Section \ref{S:3.2}, in our baseline analysis, we drop traffic segments for which traffic speeds are inconsistently recorded during the study period. If there are substantial changes in the traffic speeds of the dropped traffic segments following the implementation of traffic congestion, our baseline results would be biased. To alleviate this concern, we re-compile a superset of the traffic speed data comprising all traffic segments and re-run Eq. \ref{eq1} on this superset. Table \ref{tab:robustness check} offers the corresponding model results. We find that the re-estimated impact of the congestion tax on traffic speeds is of a relatively small magnitude (0.005) and statistically insignificant. The result aligns with our baseline results, thereby indicating that our data cleaning strategy is reasonable.

\begin{table}[!ht]
\centering
\caption{Robustness checks results: using the superset of traffic speed data}
\label{tab:robustness check}
\resizebox{0.4\textwidth}{!}{%
\begin{tabular}{@{}ll@{}}
\toprule
                              & Log(traffic speeds) \\ \midrule
TREAT×POST                    & 0.005               \\
                              & (0.006)             \\
TREAT                         & 0.011**             \\
                              & (0.005)             \\
POST                          & 0.003               \\
                              & (0.005)             \\
Weather control               & Y                   \\
Hour of day fixed effects     & Y                   \\
Day of week fixed effects     & Y                   \\
Traffic segment fixed effects & Y                   \\
No. of observations           & 209,002             \\
Adj. R-squared                & 0.713               \\ \bottomrule
\end{tabular}%
}

\parbox[t]{0.38\textwidth}{\vskip3pt{\footnotesize Notes: Standard errors clustered at the census-tract-by-day (or traffic-segment-by-day) level are provided in parenthesis. *Significant at the 10\% level, **Significant at the 5\% level, ***Significant at the 1\% level. }}
\end{table}

Third, we test whether our baseline results are robust against alternative model specifications, including the count data model setting, census-tract-year (traffic-segment-year) specific time trends, weighting samples, \xzhangadd{and additional control variables}. \xzhangadd{In our baseline analysis, we employ Eq. \ref{eq1}, i.e., the linear model with a logarithmic transformation of outcome variables, to analyze PUDOs of ridesourcing trips. As robustness checks, we alternatively utilize the negative binomial (NB) regression model, which is exclusively developed for modeling over-dispersed count data, to scrutinize the impact of the congestion tax on PUDOs of ridesourcing trips.} The first panel of Table \ref{tab:Robustness check results: alternative model specifications} reports the corresponding results. We can see that using \xzhangadd{NB} models yields estimated coefficients comparable to our baseline results. Then, we re-run Eq. \ref{eq1} with specifications of the census-tract-year (or traffic-segment-year) specific linear time trend to allow census tracts (or traffic segments) with different linear changing patterns in the outcome variables around January $6^{th}$, 2020. As shown in the second panel of Table \ref{tab:Robustness check results: alternative model specifications}, the estimated coefficients are similar to those of our baseline results. Besides, instead of weighting all census tracts or traffic segments identically as in our baseline analysis, we weight each census tract by its prior ridesourcing demand. For example, when estimating the impact of the congestion tax on PUDOs of single trips, we weight each census tract by its average PUDOs of single trips prior to the implementation of this tax. Regarding traffic segments, we weight them by their length. The third panel of Table \ref{tab:Robustness check results: alternative model specifications} reports the estimated coefficients. Again, we find our baseline results hold. \xzhangadd{In addition, we incorporate the information on weekly regular reformulated retail gasoline prices into Eq. \ref{eq1} as a control variable to see whether our baseline results may be confounded by it\footnote{\xzhangadd{The confounding effects of transit fares can be ruled out because they were stable over the study period.}}. As shown in the fourth panel of Table \ref{tab:Robustness check results: alternative model specifications}, the inclusion of the changes in fuel prices has little impact on our baseline results.}

\begin{table}[!ht]
\centering
\caption{Robustness check results: alternative model specifications}
\label{tab:Robustness check results: alternative model specifications}
\resizebox{\textwidth}{!}{%
\begin{tabular}{@{}llllll@{}}
\toprule
           & \multicolumn{4}{l}{Log(PUDOs+1)}                                             & \multirow{2}{*}{Log(traffic speeds)} \\ \cmidrule(lr){2-5}
           & Single trips & Sharing-authorized trips & Sharing-matched trips & All trips &                                      \\ \cmidrule(l){2-6} 
                 & (1)       & (2)      & (3)      & (4)       & (5)     \\ \midrule
\multicolumn{6}{l}{\textit{\xzhangadd{Negative binomial} regression model}}                    \\
TREAT×POST       & \xzhangadd{-0.157***} & \xzhangadd{0.282***} & \xzhangadd{0.210***} & \xzhangadd{-0.111***} &         \\
                 & \xzhangadd{(0.021)}   & \xzhangadd{(0.024)}  & \xzhangadd{(0.030)}  & \xzhangadd{(0.019)}   &         \\
Pseudo R-squared & \xzhangadd{0.210}     & \xzhangadd{0.266}    & \xzhangadd{0.264}    & \xzhangadd{0.210}     &         \\
\multicolumn{6}{l}{\textit{Census tract-year (or traffic segment-year) specific linear trends}}                                 \\
TREAT×POST       & -0.200*** & 0.294*** & 0.236*** & -0.145*** & 0.000       \\
                 & (0.033)   & (0.033)  & (0.038)  & (0.031)   & (0.009) \\
Adj. R-squared   & 0.936     & 0.91     & 0.895    & 0.939     & 0.625   \\
\multicolumn{6}{l}{\textit{Weighting samples}}                           \\
TREAT×POST & -0.140***    & 0.275***                 & 0.196***              & -0.096*** & 0.006                                \\
                 & (0.023)   & (0.023)  & (0.031)  & (0.021)   & (0.007) \\
Adj. R-squared   & 0.929     & 0.926    & 0.914    & 0.933     & 0.635   \\ 
\multicolumn{6}{l}{\textit{\xzhangadd{Additional control variables}}}                           \\
TREAT×POST & \xzhangadd{-0.181***}    & \xzhangadd{0.242***} & \xzhangadd{0.184***}              & \xzhangadd{-0.134***} & \xzhangadd{0.010}                                \\
                 & \xzhangadd{(0.025)}   & \xzhangadd{(0.028)}  & \xzhangadd{(0.032)}  & \xzhangadd{(0.023)}   & \xzhangadd{(0.007)} \\
Adj. R-squared   & \xzhangadd{0.935}     & \xzhangadd{0.909}   & \xzhangadd{0.894}    & \xzhangadd{0.938}     & \xzhangadd{0.625}   \\
\bottomrule
\end{tabular}%
}

\parbox[t]{0.98\textwidth}{\vskip3pt{\footnotesize Notes: The sample sizes of all five models are the same as in Table \ref{tab:Baseline results}. The coefficients of TREAT and POST are not reported for brevity. All models control for temperature, precipitation/snowfall fixed effects, hour of day fixed effects, day of week fixed effects, and census tract (or traffic segment) fixed effects. Models in the fourth panel additionally control for fuel prices. Standard errors clustered at the census-tract-by-day (or traffic-segment-by-day) level are provided in parenthesis. *Significant at the 10\% level, **Significant at the 5\% level, ***Significant at the 1\% level. }}

\end{table}

Fourth, although the parallel trend assumption holds (see Section \ref{S:4}), we cannot fully ensure that the control group provides a reasonable counterfactual scenario, given that there may be other unobserved factors affecting the control group after January $6^{th}$, 2019. To alleviate this concern, we employ regression discontinuity in time (RDiT), a widely adopted policy impact evaluation approach based on before-after comparisons \citep{hausman2017regression}, as robustness checks. RDiT is based on the assumption that the change in the outcome variable before and after the policy is only determined by the policy itself, and other observed and unobserved determinants change smoothly. By controlling for the unobserved time-varying determinants of the outcome variable using a time trend function, the causal impact of the policy can be inferred. Specifically, we employ the following sharp RDiT on the samples around January $6^{th}$, 2020:

\begin{equation}
\label{eq3}
\log \left(y_{i, t, h}\right)=\beta_{0}+\beta_{1} \operatorname{POST}_{t}+\beta_{2} f\left(X_{t}\right)+\beta_{3} \operatorname{POST}_{t} \times f\left(X_{t}\right)+\beta_{4} Z_{t, h}+\delta_{i}+\varepsilon_{i, t, h}
\end{equation}

\begin{figure}[!ht]
     \centering
     \begin{subfigure}[b]{0.475\textwidth}
         \centering
         \includegraphics[width=\textwidth]{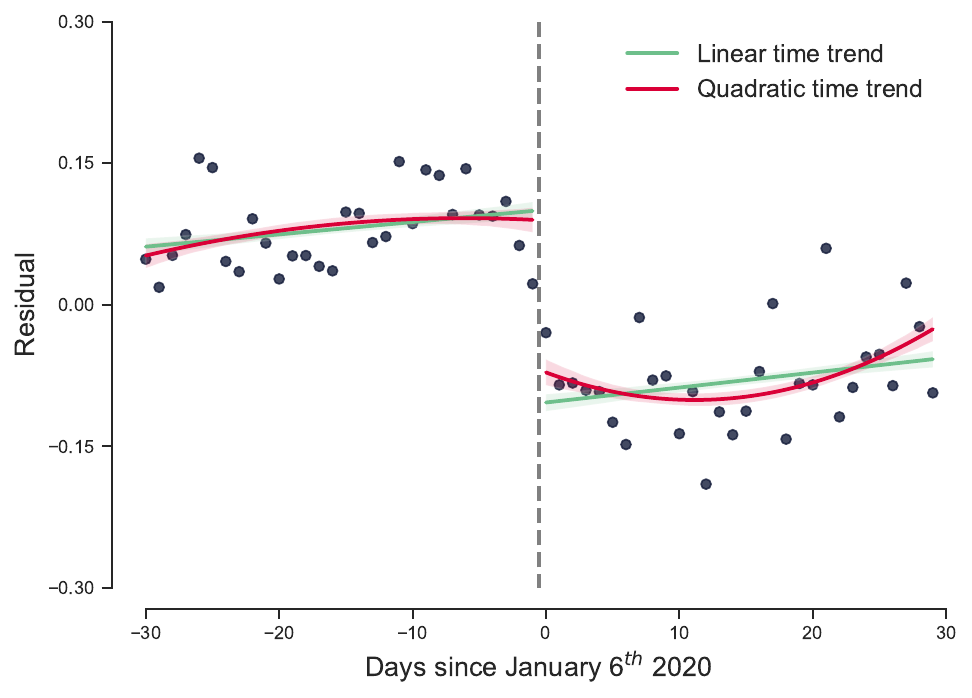}
         \caption{PUDOs of single trips}
         \label{fig:rd_PUDO of single trips}
     \end{subfigure}
     \hfill
     \begin{subfigure}[b]{0.475\textwidth}
         \centering
         \includegraphics[width=\textwidth]{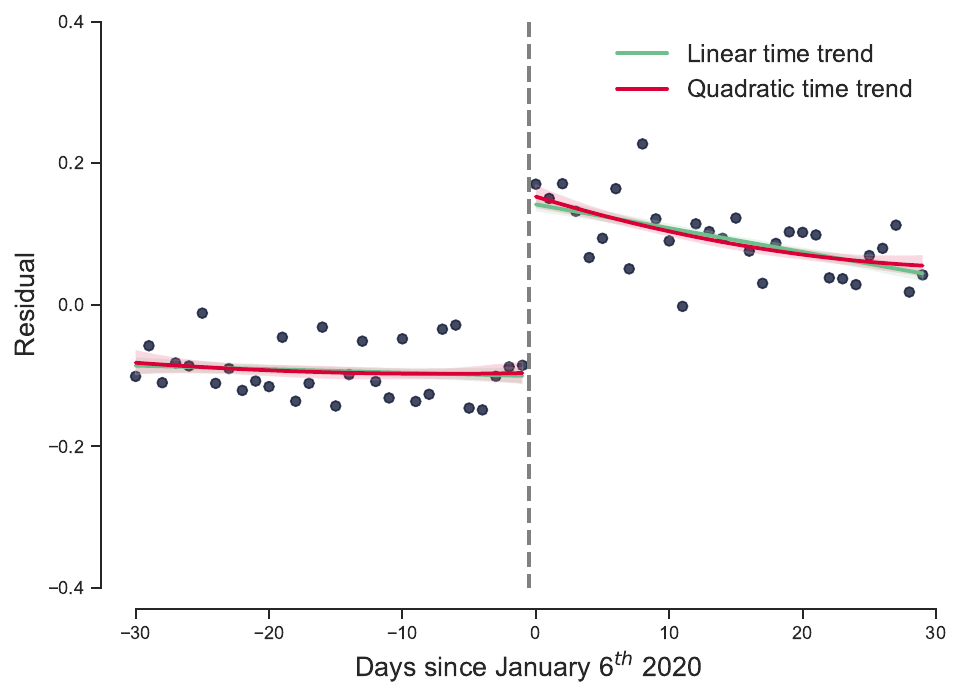}
         \caption{PUDOs of sharing-authorized trips}
         \label{fig:rd_PUDO of sharing-authorized trips}
        \end{subfigure}
    
    \begin{subfigure}[b]{0.475\textwidth}
    \centering
    \includegraphics[width=\textwidth]{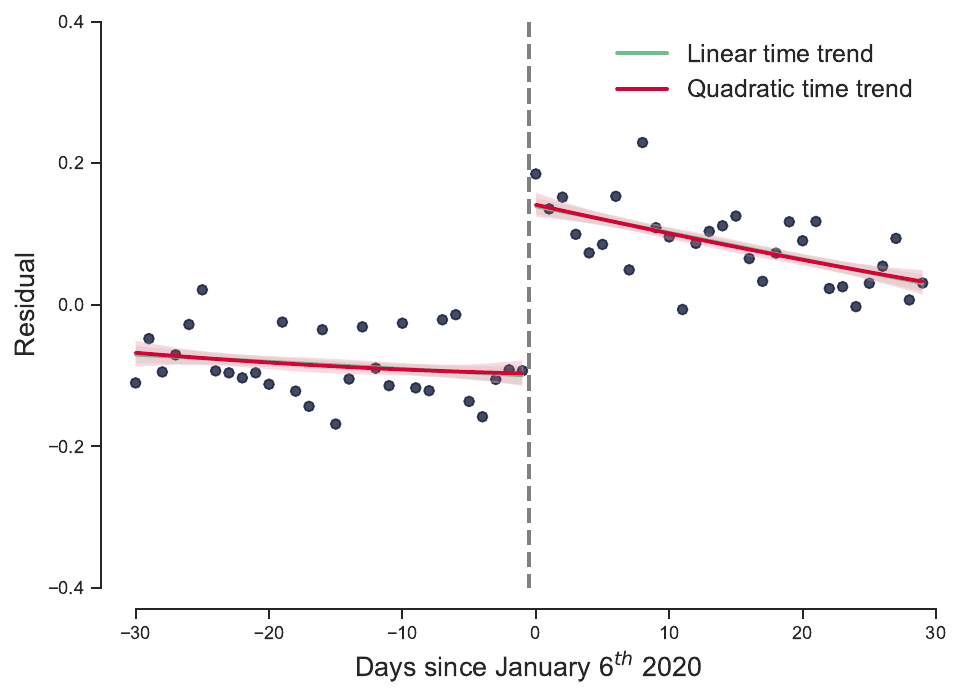}
    \caption{PUDOs of sharing-matched trips}
    \label{fig:rd_PUDO of sharing-matched trips}
     \end{subfigure}
     \hfill
     \begin{subfigure}[b]{0.475\textwidth}
    \centering
    \includegraphics[width=\textwidth]{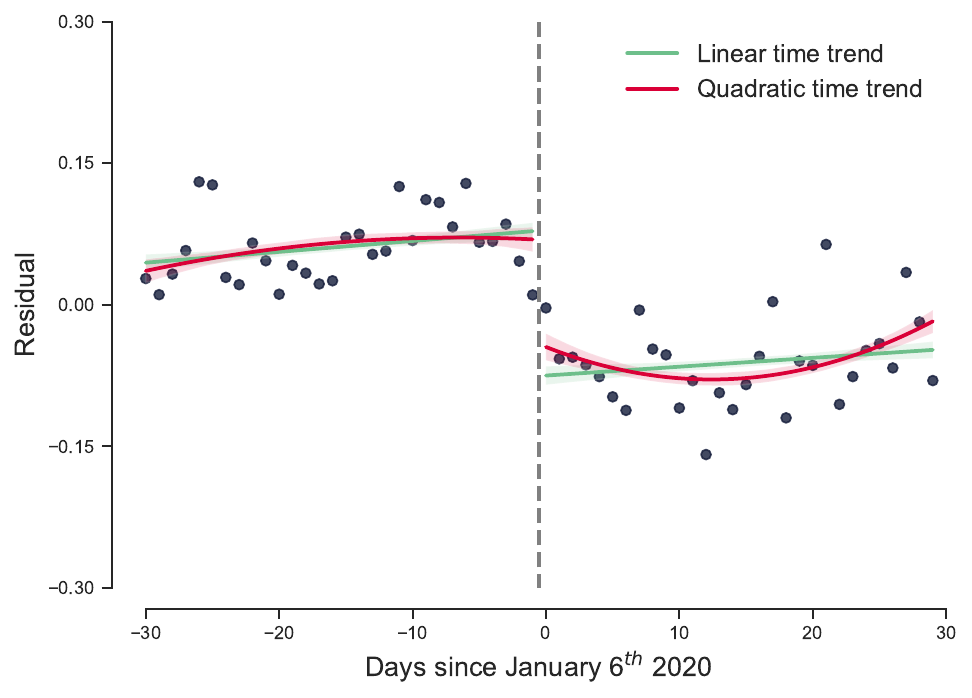}
    \caption{PUDOs of all trips}
    \label{fig:rd_PUDO of all trips}
     \end{subfigure}

    \begin{subfigure}[b]{0.475\textwidth}
    \centering
    \includegraphics[width=\textwidth]{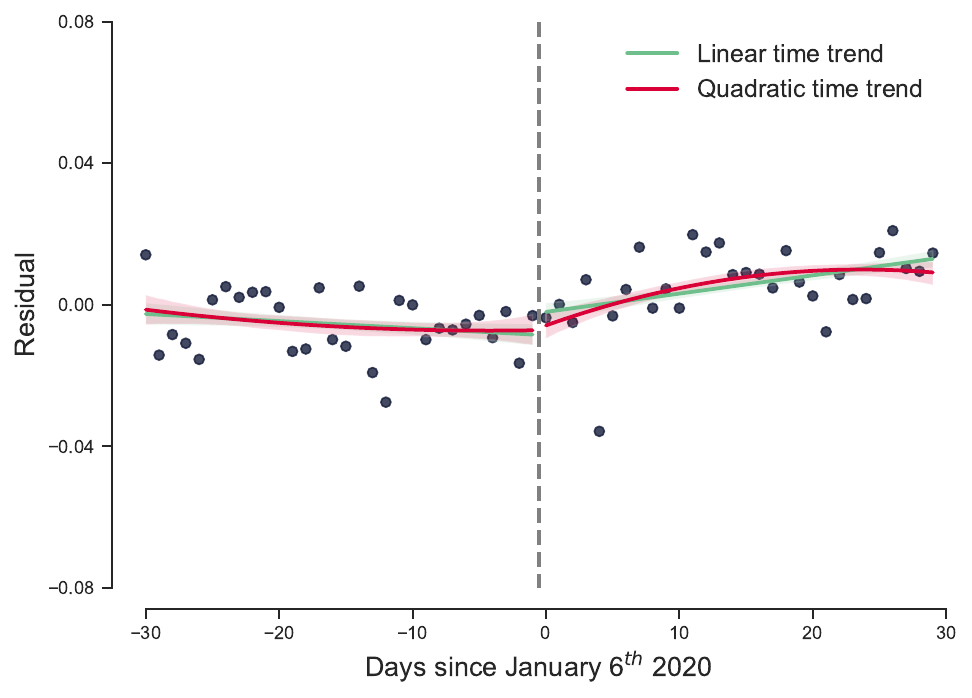}
    \caption{Traffic speeds}
    \label{fig:rd_Traffic speeds}
     \end{subfigure}
     
    \caption{Graphical illustration of regression discontinuity in time}
    \label{fig:Graphical illustration of regression discontinuity design in time}
\end{figure}

\noindent Where $X_t$ is the running variable, taking the signed number of days between day $t$ and January $6^{th}$, 2020; $f(X_t)$ is a $k$-th order polynomial function to capture the unobserved time-varying factors affecting the outcome variable; and other terms are the same as in Eq. \ref{eq1}. $\beta_{1}$ is the coefficient of interest, representing the impact of the congestion tax on ridesourcing demand or traffic congestion. Moreover, we include the interaction term of $POST_t$ and $f(X_t)$ to allow for different time trends before and after the congestion tax, which is more flexible than the global time trend. The results of RDiT models are presented in Table 6. We can observe that the impacts of the congestion tax estimated by RDiT models are largely consistent with those estimated by DID models in magnitude and direction. In addition, for ease of interpretation, Fig. \ref{fig:Graphical illustration of regression discontinuity design in time} presents the graphic evidence of the RDiT. We first estimate Eq. \ref{eq3} only with control variables and fixed effects to isolate the unexplained variation of the outcome variable as residuals. Then, we aggregate the residuals at the census-tract-by-day or traffic-segment-by-day level and plot their mean values as scatters. Finally, we separately estimate time polynomial functions on the residuals before and after the congestion tax and plot them with 95\% confidence intervals. We can see evident discontinuities for the residuals of PUDOs of ridesourcing trips. By contrast, we do not find a clear discontinuity for the residuals of traffic speeds as the time polynomial functions before and after the congestion tax are overlapped. The graphical evidence, together with the parametric estimation results, further validates the plausibility of the causal link we established in our baseline analysis.

\begin{table}[!ht]
\centering
\caption{Robustness checks results: regression discontinuity in time}
\label{tab:Robustness checks results: Regression discontinuity design in time}
\resizebox{\textwidth}{!}{%
\begin{tabular}{@{}llllll@{}}
\toprule
               & \multicolumn{4}{l}{Log(PUDOs+1)}                                             & \multirow{2}{*}{Log(traffic speeds)} \\ \cmidrule(lr){2-5}
               & Single trips & Sharing-authorized trips & Sharing-matched trips & All trips &                                      \\ \cmidrule(l){2-6} 
               & (1)          & (2)                      & (3)                   & (4)       & (5)                                  \\ \midrule
\multicolumn{6}{l}{\textit{Linear time trend}}                                                                                      \\
POST           & -0.214***    & 0.254***                 & 0.250***              & -0.162*** & 0.008                                \\
               & (0.022)      & (0.018)                  & (0.020)               & (0.021)   & (0.005)                              \\
Adj. R-squared & 0.942        & 0.896                    & 0.879                 & 0.944     & 0.600                                  \\
\multicolumn{6}{l}{\textit{Quadratic time trend}}                                                                                   \\
POST           & -0.155***    & 0.244***                 & 0.235***              & -0.109*** & 0.002                                \\
               & (0.030)      & (0.019)                  & (0.025)               & (0.028)   & (0.007)                              \\
Adj. R-squared & 0.942        & 0.896                    & 0.879                 & 0.944     & 0.600                                \\ \bottomrule
\end{tabular}%
}

\parbox[t]{0.98\textwidth}{\vskip3pt{\footnotesize Notes: $N = 27840$ for Columns 1--4 and $N = 56,640$ for Column 5. The coefficients of $f(X_t)$ and $POST \times f(X_t)$ are not reported for brevity. All models control for temperature, precipitation/snowfall fixed effects, hour of day fixed effects, day of week fixed effects, and census tract (or traffic segment) fixed effects. Standard errors clustered at the census-tract-by-day (or traffic-segment-by-day) level are provided in parenthesis. *Significant at the 10\% level, **Significant at the 5\% level, ***Significant at the 1\% level. }}

\end{table}

\newpage

\xzhangadd{Fifth, we re-compile a panel data set by combining ridesourcing trip record data from Chicago and New York City\footnote{\xzhangadd{\url{https://www.nyc.gov/site/tlc/about/tlc-trip-record-data.page}}} together and then perform a classical DID analysis to see whether our baseline results are sensitive to alternative control groups. Specifically, 47 census tracts (as shown in Fig. \ref{fig:fig_a1}) in Manhattan, south of 96$^th$ Street, act as a counterfactual to mimic what would have happened in downtown Chicago in the absence of the congestion tax. Table \ref{tab:Manhattan} presents the corresponding model results. It can be seen that the estimated coefficients are similar to those in Table \ref{tab:Baseline results}, again confirming the robustness of our baseline results.}

\begin{table}[!ht]
\centering
\caption{\xzhangadd{Robustness checks: using Manhattan, south of $96^{th}$ Street, as the control group.}}
\label{tab:Manhattan}
\resizebox{\textwidth}{!}{%
\begin{tabular}{@{}llllll@{}}
\toprule
                              & \multicolumn{4}{l}{Log(PUDOs+1)}                                              \\ 
                                \cmidrule(lr){2-5}
                              & All trips & Single trips & Sharing-authorized trips & Sharing-matched trips &                                      \\ \cmidrule(l){2-6} 
                              & (1)       & (2)          & (3)                      & (4)                 
                              \\ \midrule
TREAT×POST                    & -0.091*** & -0.119***    & 0.204***                 & 0.247***                                           \\
                              & (0.010)   & (0.010)      & (0.015)                  & (0.015)                                            \\

Weather control               & Y         & Y            & Y                        & Y                                                  \\
Hour of day fixed effects     & Y         & Y            & Y                        & Y                                                  \\
Calendar day fixed effects     & Y         & Y            & Y                        & Y                                                  \\
Census tract fixed effects    & Y         & Y            & Y                        & Y                                                  \\
No. of observations           & 72,960    & 72,960       & 72,960                   & 72,960                                             \\
Adj. R-squared                & 0.942     &   0.937        & 0.935                    & 0.916                                           \\ \bottomrule
\end{tabular}%
}
\parbox[t]{0.98\textwidth}{\vskip3pt{\footnotesize Notes: The estimated coefficients presented in this table come from a set of two-way fixed effects DID models. A total of 29 census tracts in downtown Chicago (as shown in Fig. \ref{fig:fig2}) act as the treatment group while a total of 47 census tracts (as shown in Fig. \ref{fig:fig_a1}) in Manhattan, south of $96^{th}$ Street, acts as the control group. The data used covers the time period of 30 days on either side of the implementation of the congestion tax. The weather control includes temperature and precipitation/snowfall fixed effects. Standard errors clustered at the census tract level are provided in parenthesis. *Significant at the 10\% level, **Significant at the 5\% level, ***Significant at the 1\% level.}}
\end{table}

\subsection{Heterogeneity analysis}

So far, we have identified the average impact of the congestion tax on ridesourcing demand and traffic congestion, which gives us an overview. To gain further insights, we decide to explore the heterogeneous impact of the congestion tax across different hours of the day and census tracts (or traffic segments). Moreover, we examine whether the congestion tax has heterogeneous impacts on ridesourcing demand across different travel distances.

\begin{figure}[!ht]
     \centering
     \begin{subfigure}[b]{0.475\textwidth}
         \centering
         \includegraphics[width=\textwidth]{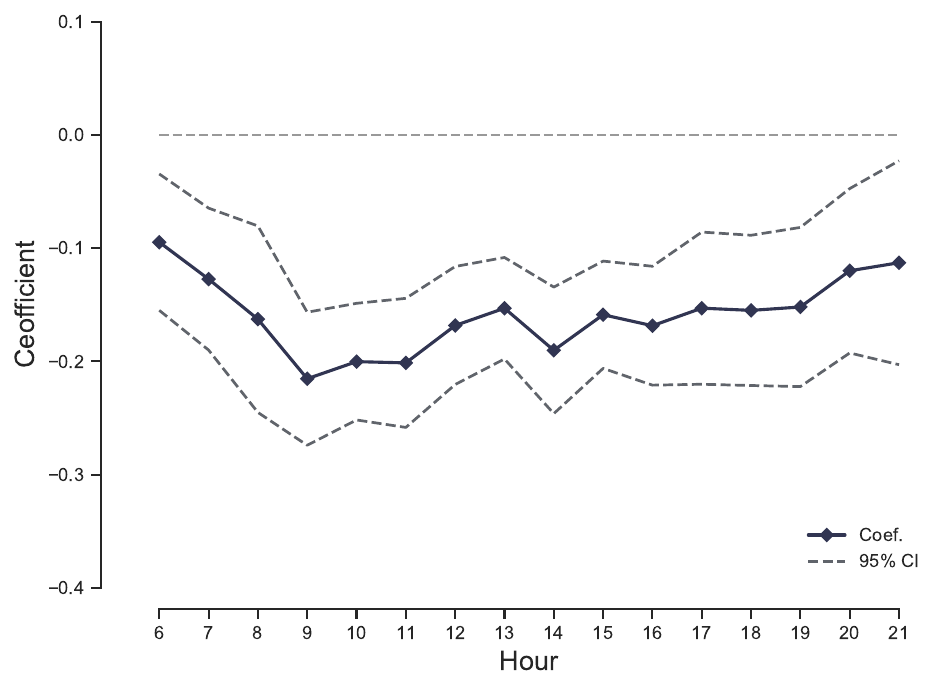}
         \caption{PUDOs of single trips}
         \label{fig:hour_PUDO of single trips}
     \end{subfigure}
     \hfill
     \begin{subfigure}[b]{0.475\textwidth}
         \centering
         \includegraphics[width=\textwidth]{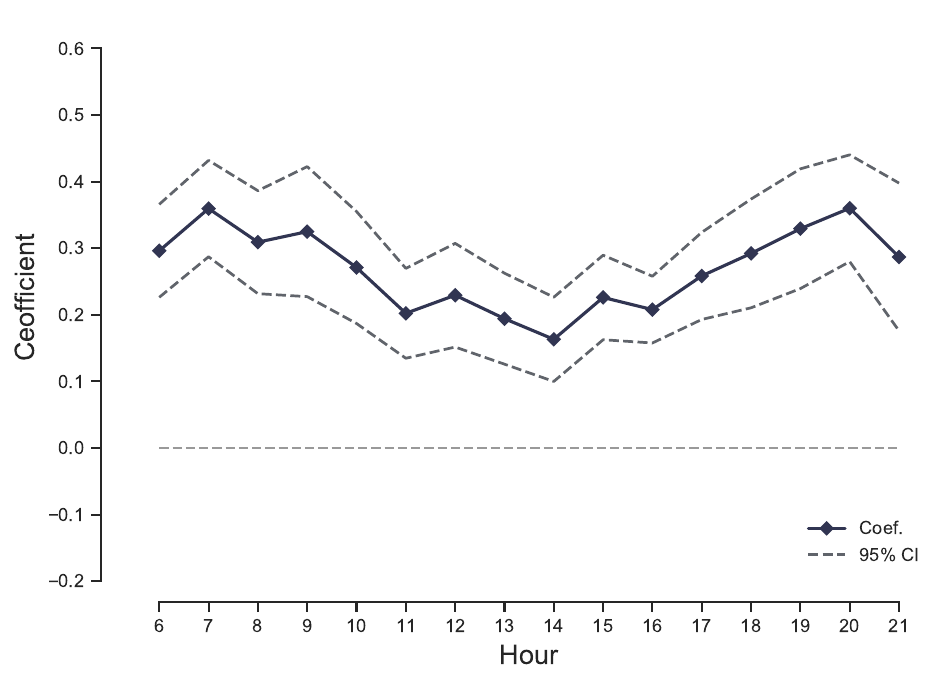}
         \caption{PUDOs of sharing-authorized trips}
         \label{fig:hour_PUDO of sharing-authorized trips}
        \end{subfigure}
    
    \begin{subfigure}[b]{0.475\textwidth}
    \centering
    \includegraphics[width=\textwidth]{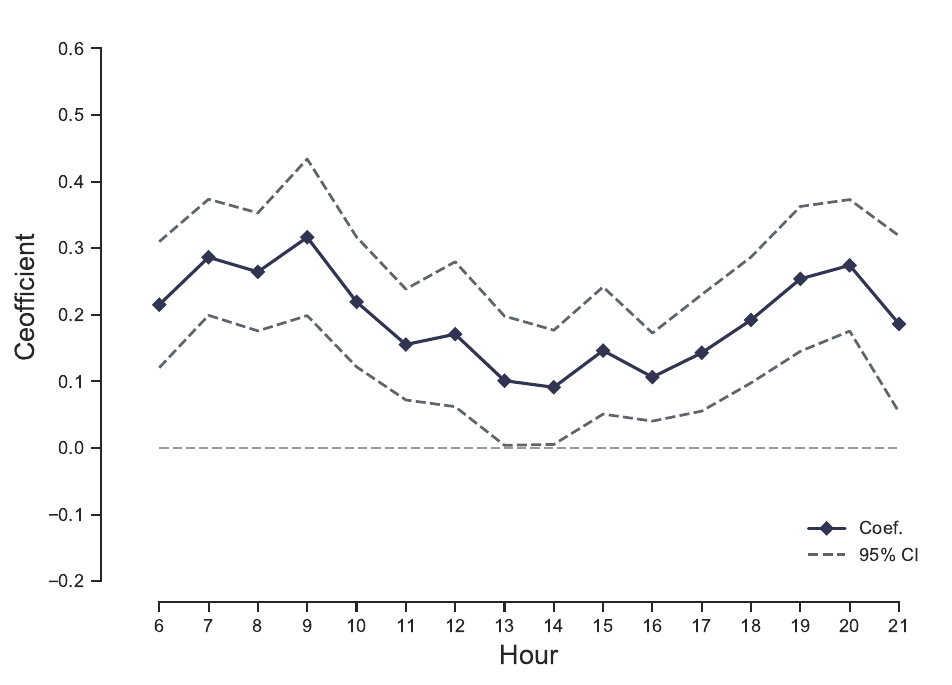}
    \caption{PUDOs of sharing-matched trips}
    \label{fig:hour_PUDO of sharing-matched trips}
     \end{subfigure}
     \hfill
     \begin{subfigure}[b]{0.475\textwidth}
    \centering
    \includegraphics[width=\textwidth]{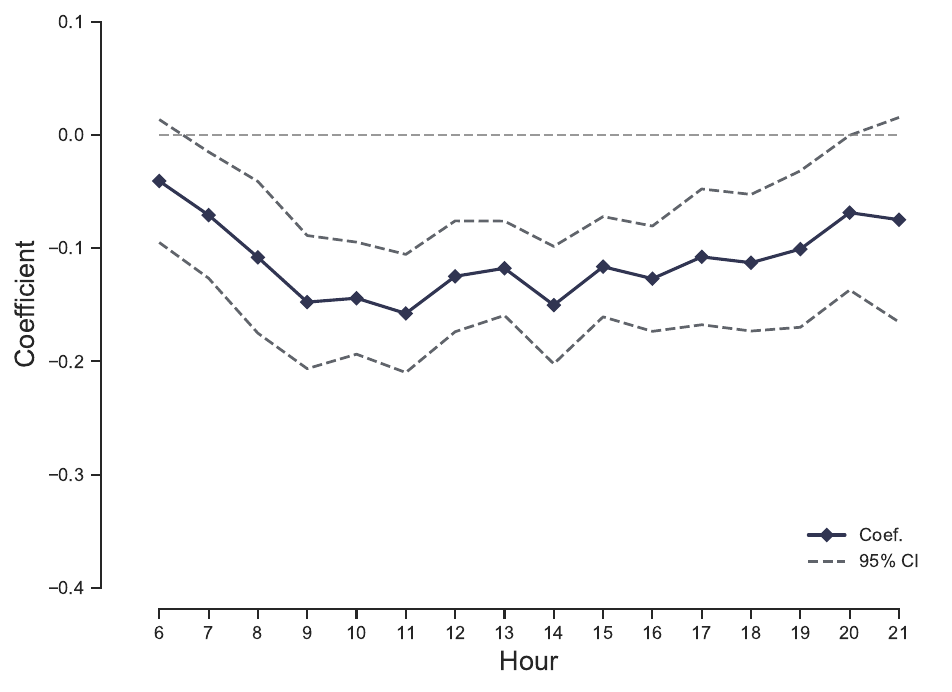}
    \caption{PUDOs of all trips}
    \label{fig:hour_PUDO of all trips}
     \end{subfigure}

    \begin{subfigure}[b]{0.475\textwidth}
    \centering
    \includegraphics[width=\textwidth]{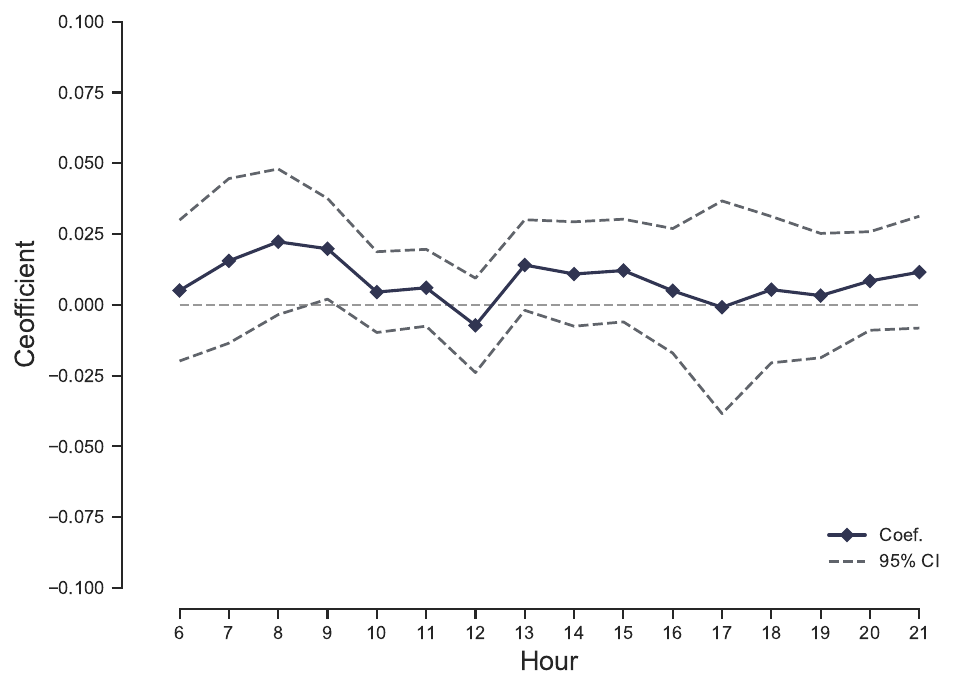}
    \caption{Traffic speeds}
    \label{fig:hour_Traffic speeds}
     \end{subfigure}
     
    \caption{Heterogeneous impacts of the congestion tax across different hours of the day}
    \label{fig:Heterogeneous hour}
\end{figure}

Fig. \ref{fig:Heterogeneous hour} depicts the estimated coefficients of the impact of the congestion tax produced by Eq. \ref{eq1} on samples of different hours of the day. For PUDOs of single trips, we observe that the estimated coefficients are all significantly negative and are of relatively small magnitudes during morning and evening hours compared to daytime hours (9 a.m. to 5 p.m.). Regarding PUDOs of sharing-authorized trips, the estimated coefficients are all significantly positive, especially in morning and evening hours (6 a.m. to 9 a.m. and 5 p.m. to 9 p.m.). And the estimated coefficients on PUDOs of sharing-matched trips also vary over different hours of the day in a similar manner, whereas their magnitudes are smaller. Meanwhile, the estimated coefficients on PUDOs of all trips exhibit a similar pattern as those on PUDOs of single trips but with smaller magnitudes. This outcome is reasonable because the impacts of the congestion tax on PUDOs of all trips consist of those on PUDOs of single and shared-authorized trips. In view of single trips accounting for a large \xzhangadd{majority} of all trips, the impacts of the congestion tax on PUDOs of all trips are heavily influenced by those on PUDOs of single trips. In a nutshell, these results indicate that the reduction impacts of the congestion tax on single trips are attenuated during morning and evening hours. On the contrary, the incentive impacts of the congestion tax on shared trips are amplified during morning and evening hours. With respect to traffic speeds, we find that nearly all estimated coefficients have positive signs but are statistically insignificant, which is in line with our baseline finding that the congestion tax has a marginal impact on relieving traffic congestion.

\begin{figure}[!ht]
     \centering
     \begin{subfigure}[b]{0.475\textwidth}
         \centering
         \includegraphics[width=\textwidth]{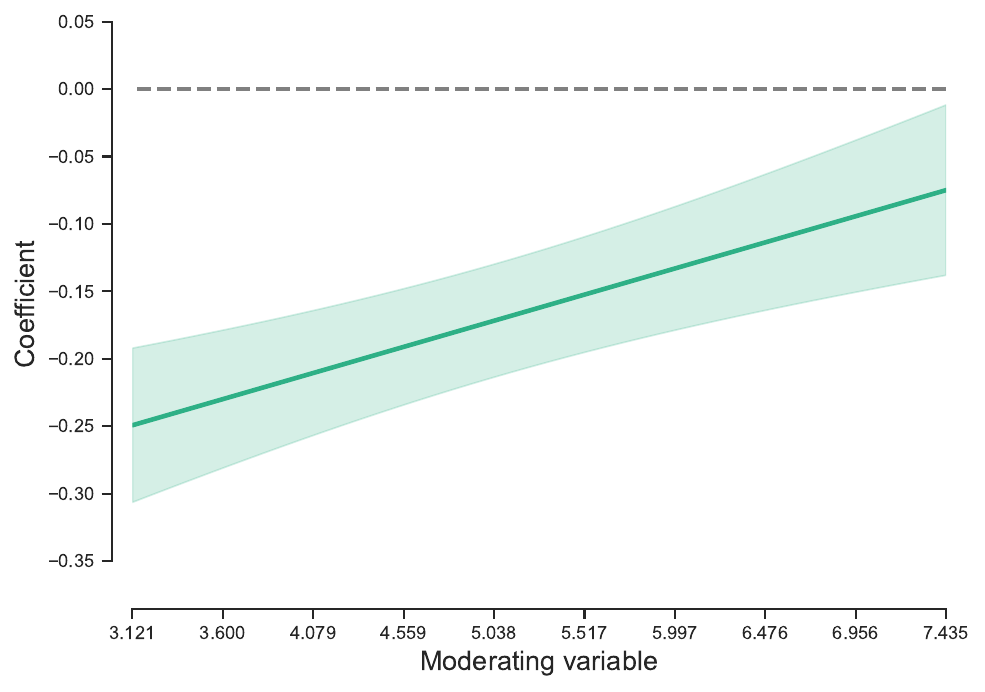}
         \caption{PUDOs of single trips}
         \label{fig:trip_PUDO of single trips}
     \end{subfigure}
     \hfill
     \begin{subfigure}[b]{0.475\textwidth}
         \centering
         \includegraphics[width=\textwidth]{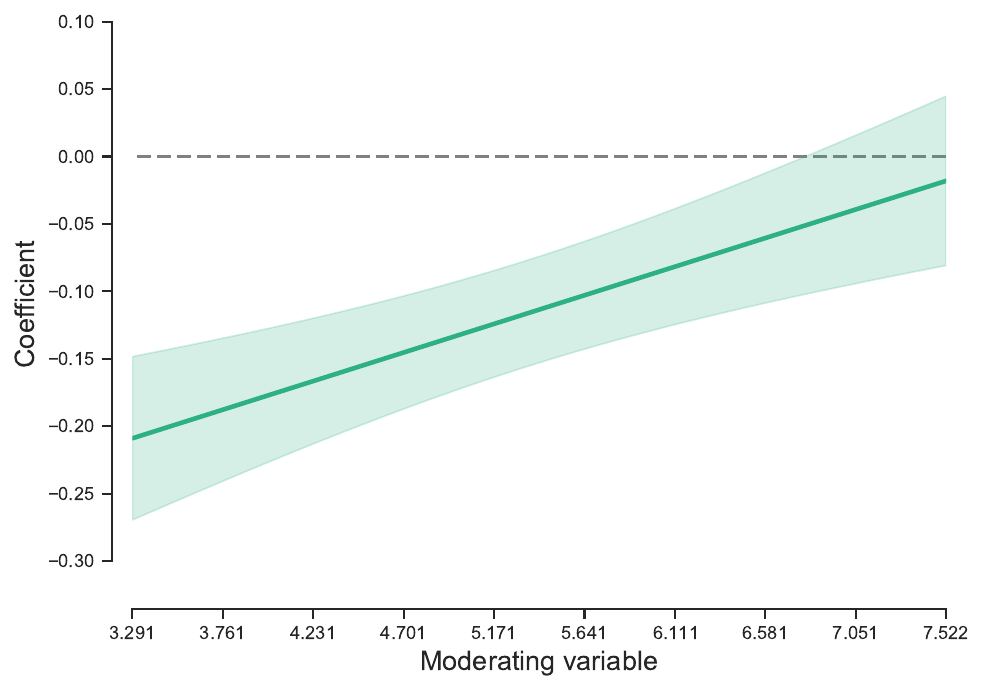}
         \caption{PUDOs of all trips}
         \label{fig:trip_PUDO of sharing-authorized trips}
        \end{subfigure}
     
    \caption{Heterogeneous impacts of the congestion tax on PUDOs of single and all trips}
    \label{fig:Heterogeneous impacts of the congestion tax on PUDO of single and all trips}
\end{figure}

In addition, we explore the heterogeneous impact of the congestion tax across different census tracts (or traffic segments) by adding triple interaction terms of $TREAT$, $POST$, and the moderating variable in Eq. \ref{eq1}. To be specific, we adopt the average level of outcome variables of census tracts (or traffic segments) prior to the implementation of the congestion tax as the moderating variable. For instance, when testing the heterogeneous impacts of the congestion tax on PUDOs of all trips across census tracts, the moderating variable is set as the census tract-level average value of PUDOs of all trips prior to the implementation of this tax. The estimated coefficient of the triple interaction terms represents whether and how the impact of the congestion tax on the demand for all trips varies across census tracts with different levels of initial demand for all trips. Table \ref{tab:Heterogeneous ct} displays the model results. We find that the estimated coefficients of $TREAT \times POST$ (as shown in the first row of Table \ref{tab:Heterogeneous ct}) are of similar magnitude, direction, and significant level to those of our baseline results. Moreover, we observe that the heterogeneous impacts only exist in models of PUDOs of single and all trips, given that the coefficients of the triple interaction term are both significant and positive. For better interpretation, we graphically visualize the two significant heterogeneous impacts in Fig. \ref{fig:Heterogeneous impacts of the congestion tax on PUDO of single and all trips}. The $y$-axis is the estimated coefficient of $TREAT \times POST$, representing the impact of the congestion tax. We can see that the coefficients are all negative across all ranges of the moderating variable. Meanwhile, the magnitude of the coefficients diminishes as the moderating variable grows. This finding suggests that the congestion tax exerts greater reduction effects in census tracts with low levels of initial ridesourcing demand than in those with high levels of initial ridesourcing demand.

\begin{table}[!ht]
\centering
\caption{Heterogeneous impacts of the congestion tax across different census tracts (or traffic segments)}
\label{tab:Heterogeneous ct}
\resizebox{\textwidth}{!}{%
\begin{tabular}{llllll}
\hline
                    & \multicolumn{4}{l}{Log(PUDOs+1)}                                             & \multirow{2}{*}{Log(traffic speeds)} \\ \cline{2-5}
                    & Single trips & Sharing-authorized trips & Sharing-matched trips & All trips &                                      \\ \cline{2-6} 
                    & (1)          & (2)                      & (3)                   & (4)       & (5)                                  \\ \hline
\textit{TREAT×POST} & -0.375***    & 0.188***                 & 0.157**               & -0.358*** & -0.04                                \\
                    & (0.056)      & (0.067)                  & (0.065)               & (0.064)   & (0.081)                              \\
\textit{\begin{tabular}[c]{@{}l@{}}TREAT×POST×\\ MODERATING VARIABLE\end{tabular}} & 0.040*** & 0.026 & 0.013 & 0.045*** & 0.016 \\
                    & (0.010)      & (0.018)                  & (0.018)               & (0.011)   & (0.028)                              \\
Adj. R-squared      & 0.935        & 0.909                    & 0.894                 & 0.944     & 0.619                                \\ \hline
\end{tabular}%
}
\parbox[t]{0.98\textwidth}{\vskip3pt{\footnotesize Notes: The sample sizes of all five models are the same as in Table \ref{tab:Baseline results}. For each column, the moderating variable is the natural logarithm of the census tract-level (or traffic segment-level) average value of the outcome variable prior to the implementation of the congestion tax. The coefficients of $TREAT$ and $POST$ are not reported for brevity. All models control for temperature, precipitation/snowfall fixed effects, hour of day fixed effects, day of week fixed effects, and census tract (or traffic segment) fixed effects. Standard errors clustered at the census-tract-by-day (or traffic-segment-by-day) level are provided in parenthesis. *Significant at the 10\% level, **Significant at the 5\% level, ***Significant at the 1\% level. }}

\end{table}

\newpage

\begin{figure}[!ht]
    \centering
    \includegraphics[width=0.5\textwidth]{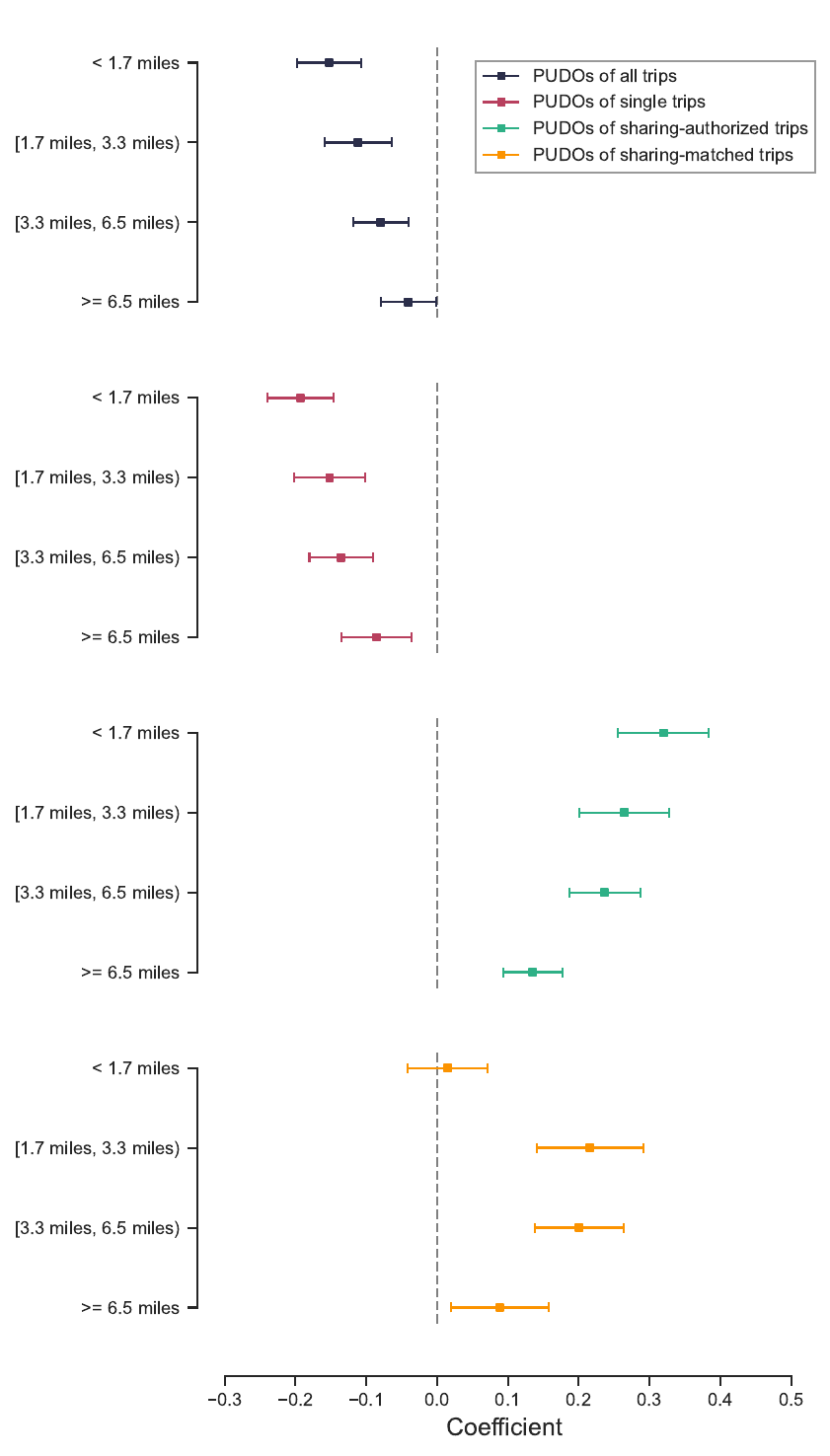}
    \caption{Heterogeneous impacts of the congestion tax across different travel distances.}
    \label{fig:fig8}
    
\end{figure}

Lastly, we re-examine Eq. \ref{eq1} on subsets of the ridesourcing trip data to investigate the heterogeneous impact of the congestion tax on ridesourcing demand across different travel distances. To this end, we first determine four intervals by using the 25th, 50th, and 75th percentiles of travel distances of all ridesourcing trips as cutoffs, which are in turn used to divide the ridesourcing trip data into four subsets with different travel distances. Then, we separately perform DID analysis on the four subsets to examine the impacts of the congestion tax on PUDOs of single, sharing-authorized, sharing-matched, and all trips. The estimated coefficients of TREAT×POST, which represent the impact of the congestion tax, are displayed in Fig. \ref{fig:fig8}. We find that for PUDOs of single and all trips, the estimated coefficients are significantly negative in four subsets. Meanwhile, the magnitude of the estimated coefficients decreases as the trip distance grows. As for PUDOs of sharing-authorized and sharing-matched trips, the estimated coefficients are all significantly positive (except for PUDOs of the sharing-matched trips with a distance of less than 1.7 miles) in four subsets. And the magnitude of the estimated coefficients decreases as the trip distance grows, too. Overall, the results indicate that the congestion tax generates a sizeable impact on reducing ridesourcing demand with short travel distances, whereas such impact is gradually attenuated as the travel distance increases. \xzhangadd{This finding is intuitive and straightforward. Since the per-trip surcharge represents a larger proportion of the fare for short trips than for long trips, short trips may be particularly sensitive to it.}

\newpage

\section{Conclusions and discussions}
\label{S:6}

Boosted by the popularization of information and communications technologies, the last decade has witnessed a tremendous expansion of ridesourcing. Notwithstanding the various promising theoretical benefits, the \xzhangadd{rapid growth} of ridesourcing is often regarded as a significant contributor to increasing traffic congestion, particularly in urban cores. How policy instruments can be leveraged to curb the excessive ridesourcing demand in urban cores has raised concerns among researchers, transportation planners, and policymakers. Although several studies have attempted to address this issue, they primarily base their conclusions on theoretical mathematical models. To date, no empirical evidence exists due to the recency of relevant policies and insufficient available data. To bridge this gap, based on recently released fine-grained ridesourcing trip and high-frequency traffic speed data, we examine the short-term impact of Chicago’s congestion taxes on ridesourcing demand and traffic congestion in the downtown area using DID, a regression-based causal inference modeling approach. To be specific, we first perform DID analysis on the full sample of ridesourcing trip and traffic speed data to identify the average impact of the congestion tax as our baseline results. We then conduct several robustness checks to verify that our baseline results are consistent and robust against different time windows, the superset of the traffic speed data used in our baseline analysis, additional control variables, alternative model specifications, alternative modeling approaches, and alternative control groups. To gain further insights, we investigate the heterogeneous impact of the congestion tax across different hours of the day, different census tracts (or traffic segments), and different travel distances by performing DID analysis on the subsets of ridesourcing trip and traffic speed data. Our empirical findings are summarized as follow.

\begin{itemize}
    \item The congestion tax has a salient impact on reducing overall ridesourcing demand associated with the downtown area, as it leads to an 10.6\% decrease in PUDOs of all trips.
    
    \item The congestion tax significantly reduces single-trip demand and promotes shared-trip demand associated with the downtown area. In particular, PUDOs of single trips decrease by 14.7\% and PUDOs of sharing-authorized and sharing-matched trips increase by 30.3\% and 21.2\%, respectively, following the implementation of the congestion tax. 
    
    \item The congestion tax has a marginal impact on alleviating traffic congestion in the downtown area, given that the estimated coefficient is only 0.008, which is equivalent to a 0.8\% increase, but statistically insignificant.
    
    \item The reduction impact of the congestion tax on single-trip demand is attenuated during morning and evening hours. By contrast, the incentive impact of the congestion tax on shared-trip demand is amplified during morning and evening hours.
    
    \item The congestion tax exerts larger reduction impacts on the ridesourcing demand of census tracts with low prior ridesourcing use than on that of census tracts with high prior ridesourcing use.
    
    \item Short-distance trips are more sensitive to the congestion tax as compared with long-distance trips.
    
\end{itemize}

There are several takeaways from our empirical findings warranting further discussions. To begin with, our empirical findings indicate that although the congestion tax notably reduces ridesourcing demand associated with the downtown area, traffic congestion is insignificantly improved. A comprehensive and in-depth deciphering of the mechanisms behind this phenomenon would require microscopic individual-level survey data that are unavailable to us, but we can draw some potential explanations from the existing literature and suggestive evidence. \xzhangadd{First, and most importantly, the finding that Chicago’s congestion tax has little or no impact on travel congestion may be related to the fact that ridesourcing trips represent a small fraction of overall trips in the system. Drawing on data from My Daily Travel survey\footnote{\xzhangadd{\url{https://www.cmap.illinois.gov/data/transportation/travel-survey}}} conducted in 2018 and 2019, we find that ridesourcing (including both solo rides and shared rides) is used for only 2.76\% of all trips associated with downtown Chicago during workday peak times, while personal automobiles are used for 31.79\% (as shown in Figure \ref{fig:fig_c1}). Given this, the effects of ridesourcing on congestion may be marginal. Therefore, the decrease in ridesourcing demand caused by the congestion tax will not significantly reduce traffic congestion. In this sense, it is reasonable to conclude that putting in policies aimed at reducing congestion caused by ridesourcing trips makes little sense without first addressing the vastly more important congestion caused by private automobile travel\footnote{\xzhangadd{\url{https://www.itf-oecd.org/app-based-ride-and-taxi-services-principles-regulation}}}.} Second, the travel mode for which ridesourcing is substituting determines whether the congestion tax, which significantly reduces ridesourcing demand, can eventually relieve traffic congestion. As the For-hire Vehicle Transportation Study published by the City of New York points out\footnote{\url{https://www1.nyc.gov/assets/operations/downloads/pdf/For-Hire-Vehicle-Transportation-Study.pdf}}, ridesourcing primarily substitutes for traditional taxis in the central business district (CBD). If this situation holds in the downtown area of Chicago, then most travelers who give up using ridesourcing following the implementation of the congestion tax may likely switch to traditional taxis, which they used to rely on. In this case, actual vehicles on the road, as well as traffic congestion, may not change substantially because the reduced ridesourcing trips would be primarily replaced by taxi trips. \xzhangadd{Third, ridesourcing driver supply is unlikely to adjust in response to reduced demand in the short run because drivers might face difficulties in substituting to alternative work arrangements. Under such circumstances, the number of ridesourcing vehicles on the road will not change much, nor will the traffic congestion.} Fourth, our measurement of traffic congestion may also be partially responsible for the insignificant impact of the congestion tax on the traffic congestion estimated by our models. As mentioned in Section \ref{S:3.2}, the traffic speed data used in our analysis mainly records the traffic condition of traffic segments that are almost arterial streets. The finding might be different if collector and local roads are taken into account.


Additionally, we find that the impact of the congestion tax on ridesourcing demand exhibits evident spatio-temporal heterogeneity. This finding implies that the demand elasticity of ridesourcing varies depending on time and space. As a result, if municipal authorities seek to maximize the demand reduction effect of per-trip surcharges for ridesourcing in the downtown area, an even charging strategy over space and time is insufficient. Instead, the spatio-temporal dynamic charging, which raises different amounts of per-trip surcharges depending on hours of the day and certain zones, needs to be considered \citep{tarduno2021congestion}. Our empirical findings also demonstrate that travel distances matter when targeting per-trip surcharges on ridesourcing. In particular, the per-trip surcharge produces only a few expected impacts (i.e., reducing single-trip demand and promoting shared-trip demand) on long-distance trips. Hence, we argue that travel distance-based, travel time-based, \xzhangadd{or travel fare-based surcharges, which are equivalent to each other,} tend to create more desirable outcomes on ridesourcing demand. Meanwhile, such charging strategies are likely to yield higher tax revenues that can be allocated to maintain and upgrade the road infrastructure and public transit system \citep{zhao2020revenue}. 

Lastly, it is a worthwhile and timely question of whether and how the impact of the congestion tax will change in the post-pandemic era or how we should frame regulation policies regarding ridesourcing in the post-pandemic era. The COVID-19 pandemic imposed a profound impact on urban mobility \citep{loa2022has}, changing the way people live and travel. There is no doubt that such an effect will generally remain in the post-pandemic era. For instance, the ridesourcing industry has witnessed disruptive changes since the start of the pandemic. On the supply side, ridesourcing companies (e.g., Uber and Lyft) encounter a shortage of drivers, given that the pandemic has expelled many drivers from this industry due to health threats\footnote{\url{https://www.cbsnews.com/chicago/news/while-demand-is-returning-a-driver-shortage-is-making-it-harder-to-get-a-rideshare-in-chicago/}}. On the demand side, travelers are less willing to use ridesourcing owing to possible exposure to strangers (i.e., the driver and ridesharers) and shared surfaces (i.e., the vehicle) \citep{loa2022has}. This situation forces us to reconsider ridesourcing itself and the role it plays in urban mobility in the post-pandemic era. Our study provides an empirical view into the short-term impact of congestion taxes on ridesourcing demand and traffic congestion in the pre-pandemic era, and we call for further effort to revisit this research topic and explore how ridesourcing should be regulated using policy instruments in the post-pandemic era.

\FloatBarrier
\newpage
\appendix

\section{}
\label{appendix_a}

\setcounter{table}{0}
\renewcommand{\thetable}{A.\arabic{table}}

\xzhangadd{To formally test the assertion that including the samples in the first week after the congestion tax will lead to biased model results, we update our panel data set accordingly and run the following model:}

\begin{equation}
\label{eqa1}
\log \left(y_{i, t, h}\right)=\beta_{0}+\beta_{1} \operatorname{POST}_{t}+\beta_{2} TREAT_{t}+\beta_{3} \text {POST\_1}_{t} \times TREAT_{t}+ \beta_{4} \text {POST\_2}_{t} \times TREAT_{t}+\beta_{5} Z_{t, h}+\delta_{i}+\varepsilon_{i, t, h}
\end{equation}

\noindent \xzhangadd{Where $POST\_1_t$ is a dummy variable with the value of one if day $t$ falls in the time period of the first week after the congestion tax and its counterparts of the prior year, and zero otherwise. $POST\_2_t$ is also a dummy variable with the value of one if day $t$ falls in the time period of the other weeks after the congestion tax and its counterparts of the prior year, and zero otherwise. Other terms are the same as in Eq. \ref{eq1}. Here we are interested in $\beta_3$, which represents the impact of the congestion tax in the first week after its enforcement.}

\xzhangadd{Table \ref{tab:tabela1} shows the corresponding model results. It can be seen that the congestion tax has a weak and insignificant impact on the demand for sharing-authorized trips in the first week after its enforcement. The reason may be that some travelers do not immediately realize that the per-trip surcharge for shared trips is much cheaper than that for single trips (\$1.25 versus \$3). As a result, they would not consider sharing their trips but simply give up using ridesourcing for travel. Therefore, including the samples in the first week after the congestion tax in our analysis will underestimate the incentive impacts of the congestion tax on shared trips.}

\begin{table}[!ht]
\centering
\caption{\xzhangadd{DID model results: including the samples in the first week after the congestion tax}}
\label{tab:tabela1}
\resizebox{\textwidth}{!}{%
\begin{tabular}{@{}llllll@{}}
\toprule
                              & \multicolumn{4}{l}{Log(PUDOs+1)}                                             & \\ \cmidrule(lr){2-5}
                              & All trips & Single trips & Sharing-authorized trips & Sharing-matched trips &                                      \\ \cmidrule(l){2-6} 
                              & (1)       & (2)          & (3)                      & (4)                               \\ \midrule
TREAT×POST\_1                    & -0.110*** & -0.123***    & -0.028                & -0.055**                             \\
                              & (0.018)   & (0.018)      & (0.022)                  & (0.026)                              \\
TREAT×POST\_2                 & -0.095*** & -0.139***    & 0.269***                 & 0.205***                             \\
                              & (0.019)   & (0.022)      & (0.024)                  & (0.031)                              \\
TREAT                         & 0.043**   & 0.186***     & -0.774***                & -0.801***                            \\
                              & (0.018)   & (0.018)      & (0.020)                  & (0.027)                              \\
POST                          & -0.040**    & -0.039*       & -0.049**                 & -0.005                             \\
                              & (0.018)   & (0.020)      & (0.017)                  & (0.024)                              \\
Weather control               & Y         & Y            & Y                        & Y                                 \\
Hour of day fixed effects     & Y         & Y            & Y                        & Y                                 \\
Day of week fixed effects     & Y         & Y            & Y                        & Y                                 \\
Census tract fixed effects    & Y         & Y            & Y                        & Y                                 \\
No. of observations           & 60,320    & 60,320       & 60,320                 & 60,320                            \\
Adj. R-squared                & 0.938     & 0.934        & 0.909                    & 0.894                                     \\ \bottomrule
\end{tabular}%
}
\parbox[t]{0.98\textwidth}{\vskip3pt{\footnotesize Notes: Standard errors clustered at the census-tract-by-day (or traffic-segment-by-day) level are provided in parenthesis. *Significant at the 10\% level, **Significant at the 5\% level, ***Significant at the 1\% level.}}
\end{table}

\section{}
\label{}

\setcounter{figure}{0}
\renewcommand{\thetable}{B.\arabic{figure}}

\begin{figure}[!ht]
    \centering
    \includegraphics[width=0.8\textwidth]{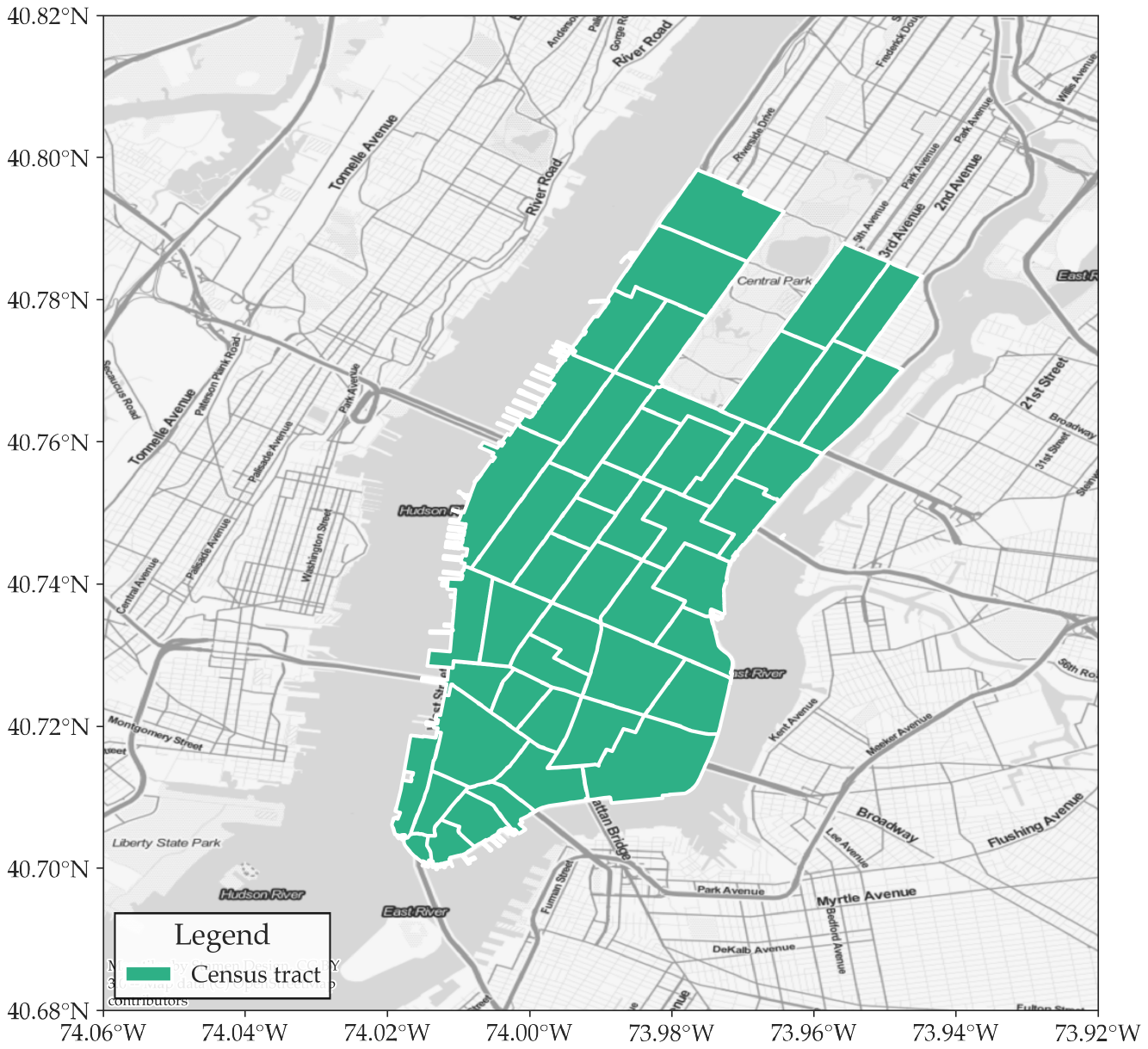}
    \caption{Census tracts in Manhattan, south of $96^{th}$ Street.}
    \label{fig:fig_a1}
    
\end{figure}
\newpage

\section{}
\label{}

\setcounter{figure}{0}
\renewcommand{\thetable}{C.\arabic{figure}}

\begin{figure}[!ht]
    \centering
    \includegraphics[width=0.8\textwidth]{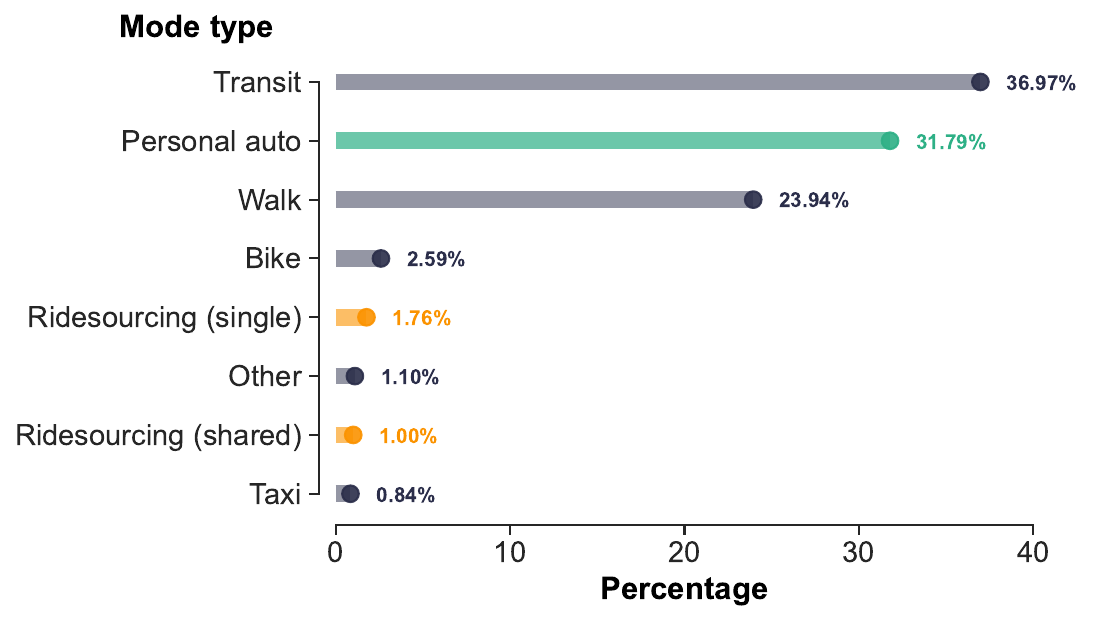}
    \caption{Mode share for daily travel associated with downtown Chicago during workday peak times}
    \label{fig:fig_c1}
    
\end{figure}

\section*{Acknowledgement}
The authors would like to thank three anonymous reviewers for their valuable comments and suggestions.




\bibliographystyle{elsarticle-harv}
\biboptions{semicolon,round,sort,authoryear}
\bibliography{sample.bib}







\end{document}